\documentclass[prb, reprint, 9pt, superscriptaddress,
notitlepage, nofootinbib, longbibliography,
floatfix]{revtex4-2}
\usepackage{natbib}
\usepackage{graphicx}
\usepackage{epsfig}
\usepackage{amsfonts} 
\usepackage{amsmath} 
\usepackage{amssymb}
\usepackage{diagbox}
\usepackage{dsfont}
\usepackage{bm}
\usepackage{braket}
\usepackage{color}
\usepackage{physics} 
\usepackage{bbm}
\usepackage{hyperref}
\usepackage{subcaption}
\usepackage{comment}

\renewcommand{\raggedright}{\leftskip=0pt \rightskip=0pt plus 0cm}
\newcommand{\be}{\begin{equation}}
	\newcommand{\ee}{\end{equation}}
\newcommand{\ba}{\begin{eqnarray}}
	\newcommand{\ea}{\end{eqnarray}}

\newsavebox{\foobox}
\newcommand{\slantbox}[1]
  {%
    \mbox
      {%
        \sbox{\foobox}{#1}%
        \hskip\wd\foobox
        \pdfsave
        \pdfsetmatrix{0.707 0.707 0 1}%
        \llap{\usebox{\foobox}}%
        \pdfrestore
      }%
  }

\definecolor{LinkColor}{rgb}{0,0,1}
\hypersetup{
	colorlinks=true,
	citecolor=LinkColor,
	linkcolor=LinkColor,
	urlcolor=LinkColor
}

\newcommand{\Lu}[1]{ { \color{blue} \footnotesize (\textsf{Lu}) \textsf{\textsl{#1}} } }

\definecolor{gr}{rgb}{0,0,0}

\begin{document}
	
	%------------------------------------------------------------
\title{Measurement as a shortcut to long-range entangled quantum matter}
\author{Tsung-Cheng Lu}
\affiliation{Perimeter Institute for Theoretical Physics, Waterloo, Ontario N2L 2Y5, Canada}

\author{Leonardo A. Lessa}
\affiliation{Perimeter Institute for Theoretical Physics, Waterloo, Ontario N2L 2Y5, Canada}
\affiliation{Department of Physics and Astronomy, University of Waterloo, Waterloo, Ontario N2L 3G1, Canada}

\author{Isaac H. Kim}
\affiliation{Department of Computer Science, University of California, Davis, CA 95616, USA}

\author{Timothy H. Hsieh}
\affiliation{Perimeter Institute for Theoretical Physics, Waterloo, Ontario N2L 2Y5, Canada}

	\begin{abstract}
The preparation of long-range entangled states using unitary circuits is limited by Lieb-Robinson bounds, but circuits with projective measurements and feedback (``adaptive circuits'') can evade such restrictions.  We introduce three classes of local adaptive circuits that enable low-depth preparation of long-range entangled quantum matter characterized by gapped topological orders and conformal field theories (CFTs). The three classes are inspired by distinct physical insights, including tensor-network constructions, multiscale entanglement renormalization ansatz (MERA), and parton constructions. A large class of topological orders, including chiral topological order, can be prepared in constant depth or time, and one-dimensional CFT states and non-abelian topological orders with both solvable and non-solvable groups can be prepared in depth scaling logarithmically with system size. We also build on a recently discovered correspondence between symmetry-protected topological phases and long-range entanglement to derive efficient protocols for preparing symmetry-enriched topological (SET) order and arbitrary CSS (Calderbank-Shor-Steane) codes.  Our work illustrates the practical and conceptual versatility of measurement for state preparation.  

	\end{abstract}
	
	\maketitle

	%------------------------------------------------------------
	
	{
		\hypersetup{linkcolor=black}
		\tableofcontents
	}

	%\newpage
	
	%=================================================================================================================================
	% \input{sec_intro}
	
	%\newpage\newpage
	\section{Introduction}

Long-range entangled (LRE) and short-range entangled (SRE) states belong to distinct  quantum phases as they cannot be connected using finite-depth unitary circuits. Intriguingly, this dichotomy can vanish if one allows for non-unitary evolution incorporating measurements. This is illustrated by the  constant-time preparation of Greenberger-Horne-Zeilinger (GHZ) cat states \cite{Raussendorf_2001_ghz} as well as toric code ground states \cite{3d_cluster_state_2005,cirac_2008_optical,cirac_2021_locc} via certain measurement protocols. More recently, it was found that performing measurements on SRE states in certain symmetry-protected-topological (SPT) phases \cite{spt_1d_2011,spt_2011} can lead to a large class of LRE states in topologically ordered phases \cite{ashvin_2021_measurement,verresen2021_measurement_cold_atom}. All these findings suggest that, in contrast to the conventional understanding that measurement simply disentangles quantum particles, performing measurement can in fact serve as a powerful tool for efficiently preparing LRE quantum matter. 

It is of great interest to explore the utility and limitations of measurements in  quantum many-body systems from various perspectives. First, certain LRE systems with topological order are ideal platforms for realizing topological quantum computation \cite{kitaev2003fault}, which allows for robust information storage and operations immune to local noise. It is, therefore, desirable to develop a scheme that can realize these LRE systems efficiently, and this endeavor is timely as mid-circuit measurements are now a capability in several experimental platforms~\cite{Chen2021,Ryan-Anderson2021}. Second, measurements can potentially lead to novel non-equilibrium states of matter that do not naturally arise from unitary evolution alone. One notable example of recent interest is hybrid quantum circuits \cite{fisher_2018_mit,amos_mit_2019,nahum_mit_2019}, where the interplay between unitary gates and projective measurements can lead to states with fascinating entanglement structures including quantum critical points that cannot be realized in equilibrium. However, a challenge is that such properties are features of pure state trajectories defined by the measurement outcomes, and thus experimentally probing such features generally requires post-selection of the outcomes.  This motivates the consideration of adaptive dynamics in which the measurement outcomes are fed forward to inform the choice of subsequent operations; then it is possible that different trajectories can converge and obviate the need for postselection.   

\begin{figure}
	\centering
    \begin{subfigure}{0.47\textwidth}
        \includegraphics[width=\textwidth]{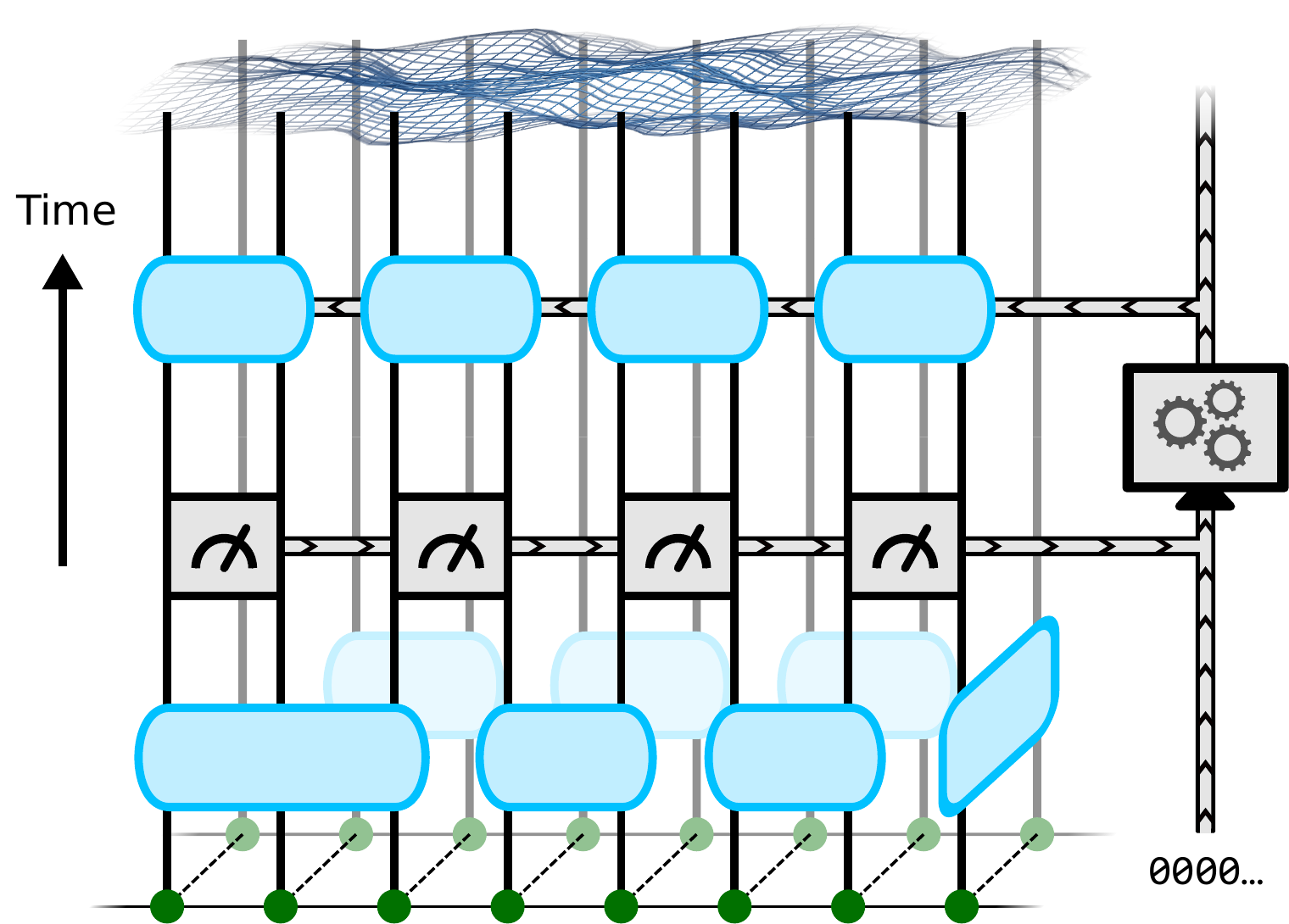}
    \end{subfigure}
    \caption{A schematic of a local adaptive circuit for preparing a long-range entangled state. The blue boxes and the gray boxes represent local unitary gates and local measurements, respectively. These operations may involve a finite number of ancilla copies (light green) of the system (green), and the measurement outcomes can be classically processed and communicated to every future component of the circuit.}
	\label{fig:AdaptiveCircuit}  
\end{figure}

In this work, we explore the utility of measurements as a versatile tool for constructing various long-range entangled states, including both chiral and non-abelian topological orders as well as 1d gapless critical states characterized by 1+1D conformal field theories. The fast preparation of target states is made possible using  ``local adaptive quantum circuits'', which we define to consist of local unitary gates, local projective measurements, classical information processing, and finite overhead of ancillae, i.e., at most a constant number of copies of the original system (see Fig. \ref{fig:AdaptiveCircuit}). The adjective ``adaptive'' means that the measurement outcomes are recorded as classical data that can be processed and communicated non-locally, and the choice of a local unitary gate after measurements depends on this global classical data.

The target state  is thus prepared without needing to post-select on particular measurement outcomes. In this work, we introduce three distinct classes  of local adaptive circuits, each of which relies on completely different physical insights and therefore, possesses different merits which we highlight below.  We also build on \cite{ashvin_2021_measurement,verresen2021_measurement_cold_atom} to derive adaptive circuits preparing symmetry-enriched topological order and arbitrary CSS (Calderbank-Shor-Steane) codes.  Each approach is self-contained and can be understood independently.

The first class of adaptive circuits (Sec.\ref{sec:tensor}) is inspired by the tensor network construction of quantum states (see e.g. Ref.\cite{cirac_2021_tn_review} for a review) and serves as a practical, modular approach for preparing many topologically ordered states in constant depth. After preparing copies of small entangled resource states of constant sizes, performing two-body measurements ``glues'' these resource states together, forming a long-range entangled many-body state. This construction applies to the abelian topological order of quantum double models \cite{kitaev2003fault}, as well as the double semion ground state \cite{wen_string_net2005}, which is a twisted quantum double.  Notably, it also allows one to prepare a perturbed topologically ordered state that remains in the same phase as a fixed-point wave function. Our protocol is similar in spirit to fusion-based quantum computation \cite{fusion_based_2021} and ``quantum legos'' \cite{cao_lego_2021}.  This modular approach involving preparing many small states in parallel is particularly well-suited for photonics and trapped ion platforms.

The second class of adaptive circuits (Sec.\ref{sec:mera})  is inspired by ``multiscale entanglement renormalization ansatz'' (MERA) \cite{vidal_2007_mera,vidal_2008_mera}, a framework for real-space renormalization group (RG) transformations where short-distance entanglement is systematically removed. MERA can also be viewed as a $O(\log L)$-depth (spatially non-local) unitary circuit that realizes  microscopic models of size $L$ by reversing the direction of RG-flow. While the depth of a MERA circuit is relatively low, it requires non-local unitary gates to generate entanglement at various length scales. We introduce a class of local adaptive circuits to implement a MERA circuit, where any non-local gate is realized with local measurements and unitaries. This immediately leads to two notable classes of target states achievable with $O(\log L )$-depth local adaptive circuits: (i) gapped topological orders in quantum double models \cite{kitaev2003fault} of any finite group and Levin-Wen string-net models \cite{wen_string_net2005} (ii) gapless states characterized by conformal field theories (CFTs). A CFT state of size $L$ in one space dimension exhibits an $O(\log L)$ entanglement scaling  \cite{wilczek_1994,kitave_2003_entanglement,vidal_2004_entanglement,calabrese_2004,cardy_2009_cft}, and we show that preparing such a state using any local adaptive circuit requires a depth that at least scales with $O(\log L)$. Our proposal, therefore, provides an optimal circuit structure for preparing 1d CFT states.

The third class of  adaptive circuits (Sec.\ref{sec:partons})  is inspired by parton constructions, an approach widely used to characterize various topological orders (see e.g. \cite{wen2004quantum}). Notably, this class of circuits enables the preparation of a chiral non-abelian topological order in finite time. We introduce the circuit in the context of the Kitaev honeycomb model \cite{kitaev_2006}, where each qubit is fractionalized into four Majorana partons subject to a local constraint. The model can be mapped to a system of free Majorana fermions coupled to static $\mathbb{Z}_2$ gauge fields represented by Majorana dimers, and the physical state of the honeycomb model can be obtained by projecting to the physical subspace consistent with the local gauge constraints. In certain parameter regimes,  a time-reversal symmetry-breaking perturbation leads to a topological band structure for the Majorana fermions, and after projecting to the physical subspace of the spin model, one obtains the chiral topological order of Ising anyon theory. This parton-based solution motivates the construction of the following adaptive circuit: starting from a tensor product state of the free Majorana fermion state and Majorana dimers arranged on a honeycomb lattice, one can simultaneously measure certain Majorana operators to enforce local gauge constraints, thereby preparing the target chiral topological order. Crucially, the (abelian)   $\mathbb{Z}_2$ gauge structure allows us to enforce the correct gauge constraints by applying a finite-depth quantum circuit, despite the existence of non-abelian topological order. While we only discuss the parton-based adaptive circuit in the context of the Kitaev honeycomb model, we expect broad applicability of this method for preparing various topological phases of matter.

Finally, in Sec.\ref{sec:disentangling}, we build on recent works \cite{ashvin_2021_measurement,verresen2021_measurement_cold_atom}, which point out the emergence of long-range order by measuring certain SRE states with SPT order. We adopt a wave-function perspective in which certain SRE states can be understood as two species of fluctuating topological objects with a non-trivial braiding phase \cite{dwSPT}. These  topological objects are domain walls of classical models (e.g they can be point-like objects on the boundary of open strings as in 1d Ising model or loop-like objects on the boundary of open membranes as in 2d Ising model). Measuring one species ``disentangles'' the two species:   the measured species is projected to a particular domain-wall product state, and the unmeasured species remains in a superposition of all possible domain-wall configurations, which constitutes long-range order. This wave-function perspective provides the following three advances: (i) It provides a simple physical picture demonstrating that measuring an SPT with $\mathbb{Z}_2  \times\mathbb{Z}_2$ symmetry leads to a $\mathbb{Z}_2$ long-range order, a result that has been discussed in  Ref.\cite{ashvin_2021_measurement,verresen2021_measurement_cold_atom,erez_2022_duality}. (ii) Considering domain walls that result from certain classical models without spatial locality, one can construct a finite-depth adaptive circuit that prepares any CSS code \cite{css_Steane,css_Shor}. While it has been known that any CSS code can be obtained by measuring certain SRE states \cite{stace_2016_css}, we describe the construction in terms of chain complex, which provides a succinct formalism that is particularly suitable for preparing certain quantum low-density parity-check (LDPC) codes of recent interest (see Ref.\cite{niko_2021_ldpc_review} for review and recent progress). (iii) By decorating  fluctuating topological objects with lower-dimensional SPTs, one can easily prepare various symmetry-enriched topological (SET) orders \cite{wang_2019_set}, where the deconfined anyonic excitations may exhibit a further symmetry fractionalization, as they live on the boundary of SPTs.

\section{Adaptive circuits via tensor networks}\label{sec:tensor}
Here we present a class of finite-depth adaptive circuits that  prepares a large class of topologically ordered states by exploiting their tensor-network representations. This is applicable to abelian quantum double models \cite{kitaev2003fault} and a twisted quantum double -- the double semion model \cite{wen_string_net2005}. In addition, the adaptive circuits can also prepare non-fixed-point states with topological order. Below we will illustrate the main idea using the 2d toric code as an example.
	
\subsection{An example: 2d toric code}
Consider a 2d lattice with every link accommodating a qubit. The 2d toric code Hamiltonian is $- \sum_{v}  \prod_{ l \vert v\in \partial l     }Z_l  -  \sum_{ p  }   \prod_{ l |  l \in \partial p   }X_l$, where the first term is a product of four Pauli-Zs acting on links emanating from a given vertex $v$, and the second term is a product of four Pauli-Xs acting on links around the boundary of a plaquette $p$. With periodic boundary conditions for both two spatial directions (i.e. the lattice is defined on a 2-torus), we consider a (un-normalized) ground state as a superposition of all possible  loops $\mathcal{C}$, including both the contractible ones and non-contractible ones:
\begin{equation}\label{eq:toric}
\ket{ \mathcal{T} } = \sum_{ \mathcal{C}} \ket{\mathcal{C}}.
\end{equation}
The states $\ket{\mathcal{C}}$ form  the computational (Pauli-Z) basis of loop configurations, where qubits take  the value $1$ (down spin), $0$ (up spin) depending on whether they  belong to a loop or not. In other words, the set of states $\ket{\mathcal{C}}$ is all possible spin configurations with an even number of down spins around each vertex.

To introduce our protocol, we here provide a short review of the tensor-network  construction for the toric code \cite{wen_tnrg_2008,tn_string_net_wen_2009}. To start, one defines a  three-leg tensor $g^s_{i,j}$ on every link:
\begin{equation}\label{eq:g_tensor}
g_{ij}^s= \delta_{s,i} \delta_{s,j},
\end{equation}
where $s \in \{0,1\}$ labels the state of a physical qubit and $i, j \in \{0,1\}$ correspond to virtual legs. Essentially, $g$ is a projector enforcing that the physical leg and two virtual leg variables take the same value.
In addition,  on every vertex one defines a four-leg tensor $T_{\alpha \beta\gamma \delta}$
	
	\begin{equation}\label{eq:T_2dtoric_constraint}
		T_{\alpha \beta\gamma \delta} =   \begin{cases}
			1 \text{ if   } ~\alpha +  \beta + \gamma + \delta  =0 \text{ mod } 2\\
			0 \text{ if   } ~\alpha +  \beta + \gamma + \delta  =1 \text{ mod } 2. 
		\end{cases}
	\end{equation} 
	
\begin{figure}
		\centering
		\begin{subfigure}{0.46\textwidth}
			\includegraphics[width=\textwidth]{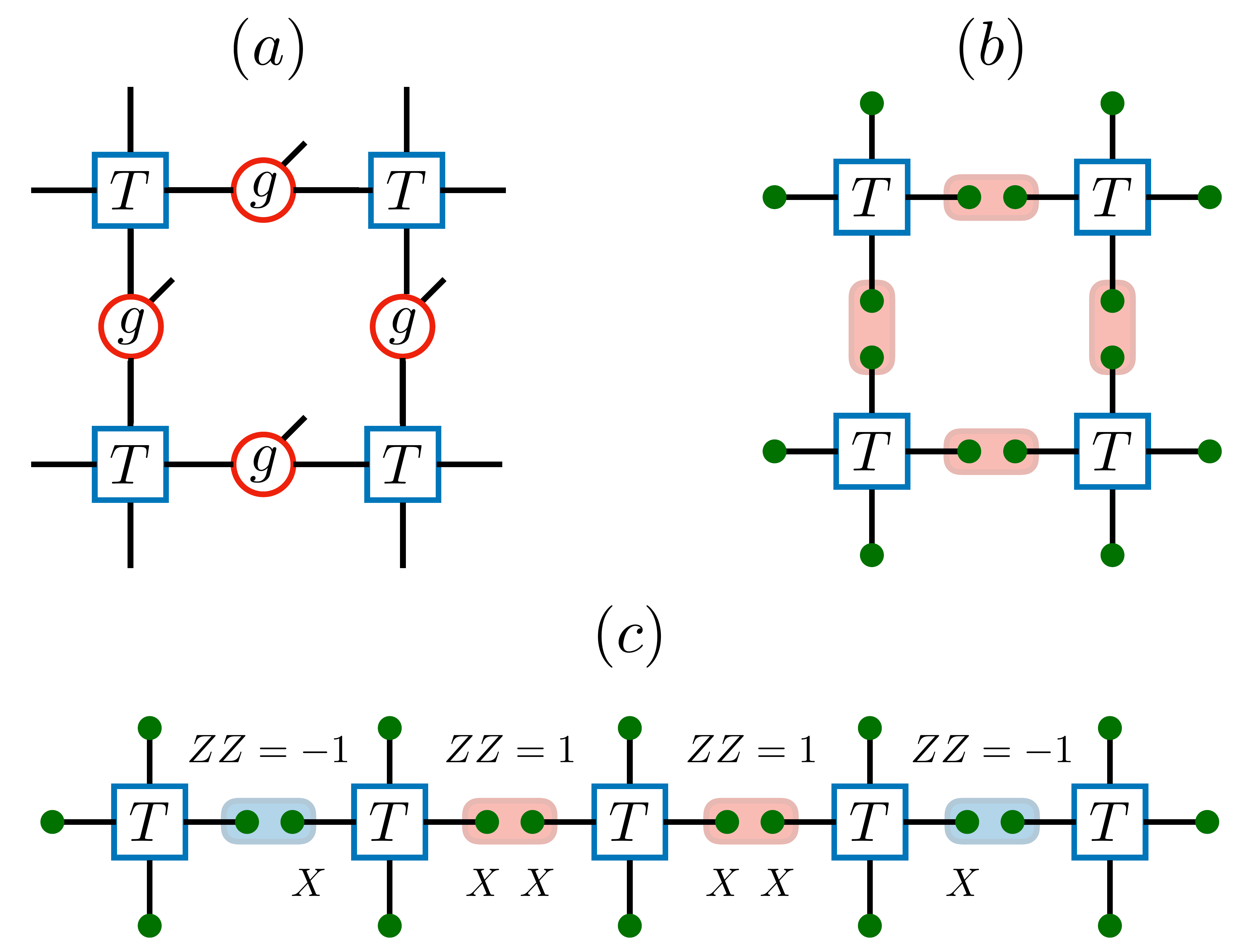}
		\end{subfigure}
		\caption{Adaptive circuits via tensor-network representation -- ground state of 2d toric code as an example: (a) Tensor network for toric code constructed from the tensor $g$ (Eq.\ref{eq:g_tensor}) and tensor $T$ (Eq.\ref{eq:T_2dtoric_constraint}). (b) One first simultaneously prepares copies of a 4-qubit vertex state, where each vertex state is defined from tensor $T$. Measuring $ZZ$ operators  for two spins on the same link  with the outcomes $ZZ=1$ (represented by red rectangles)  ``fuses'' them to a single qubit, thereby exactly preparing the toric code state. (c) Any measurement errors ($ZZ=-1$, represented by blue rectangles) occur in pairs and can be corrected in parallel by applying  string operators of Pauli-Xs that connect them.} 
		\label{fig:1}  
	\end{figure}

	It follows that the toric code ground state can be expressed via the tensors $g$ and $T$ (see Fig.\ref{fig:1}a):
	\begin{equation}
\ket{ \mathcal{T} }   =  \sum_{\{s_l\}}    \text{tTr}  \left[   \otimes_v T \otimes_l g^{s_l}   \right]  \ket{ \{s_l\} }
	\end{equation}
	with tTr denoting a tensor trace that contracts all the virtual degrees of freedom. It is easy to see that such a representation exactly represents the superposition of  loops as the $g$ tensor conveys the physical spin to virtual spins and the $T$ tensor enforces the rule that there must be an even number of occupied links emanating from vertices, indicating that any spin configurations must form loops.

	Exploiting this representation, we now introduce the  protocol for preparing the toric code ground state from SRE states via measurements. To start, we employ the  $T$ tensor to define a state of four spins associated with a vertex, i.e. $\ket{T}_v = \sum_{\alpha,\beta,\gamma,\delta}T_{\alpha \beta\gamma \delta}    \ket{ \alpha,\beta,\gamma,\delta }$ so that it is a superposition of product states for four spins subjected to the constraint $ZZZZ=1$ (every allowed state must have an even number of strings ($Z=-1$) meeting at the vertex $v$).  $\ket{T}_v$ is a stabilizer state specified by the $ZZZZ$ stabilizer and three other independent $XX$ stabilizers acting on any two qubits; it is a GHZ state in the $X$-basis. Defining an SRE resource state as $\otimes_v \ket{ T}_v  $, i.e. the tensor product of $\ket{T}_v$ among all vertices, we measure the two-body $ZZ$ operators for  two spins associated with two vertices sharing the same link (see Fig.\ref{fig:1}b). When the measurement outcomes are $ZZ=1$ on all links, measurements effectively ``fuse'' two  spins  on the same link to a single spin, in the sense that they must take the same value in the computational basis. Due to the constraint of only even number of strings ($Z=-1$) on each vertex, the post-measurement state is a superposition of closed strings (i.e. loops), thereby realizing the exact topological order of $\mathbb{Z}_2$ toric code. Our protocol relies on encoding the hard constraint on each vertex state $\ket{T}_v$, and this is in a spirit similar to the construction of classical vertex models, where vertex constraints are enforced and the lattice is tiled with mismatches of bonds that need to be removed (see e.g. Ref.\cite{color_model_castelnovo_2018} Appendix.B for details).

		Formally, the post-measurement  state is 
	\begin{equation}
		\ket{\psi}	 = \left[ \prod_l  \frac{1+  \prod_{ v \in \partial l }Z_{l,v}   }{2} \right] \otimes_v \ket{ T}_v, 
	\end{equation}
	where  $\prod_{ v \in \partial l }Z_{l,v} $ is a  product of two Pauli-Zs on two spins associated with the two neighboring vertices connected by the link $l$.  One can also write down a parent Hamiltonian $H$ for which $\ket{\psi}$ is a ground state

	\begin{equation}\label{eq:2dtoric_post_measurement}
		H= - \sum_{ v }  \prod_{ l |  v \in \partial l  } Z_{l,v} -  \sum_{p} \prod_{l | l \in \partial p  } X_{l,v}  - \sum_{l} \prod_{ v \in \partial l }Z_{l,v}. 
	\end{equation}

Due to the constraint  $\prod_{ v \in \partial l }Z_{l,v} = 1$, the 4-dimensional Hilbert space of two spins on every link is restricted  to 2 dimensions, and one can define an effective Pauli-X and Pauli-Z acting on this subspace: $\tilde{Z}_l =Z_{ l,v} = Z_{ l,v'}$ and $\tilde{X}_l  =\prod_{ v \in \partial l }X_{l,v} =  X_{ l,v}  X_{ l,v'} $. It follows that in the basis of $\tilde{Z}$, the state $\ket{\psi}$ is a superposition of loop configurations as in Eq.\ref{eq:toric}.

Upon measurement, we may obtain $ZZ=-1$ (as opposed to $ZZ=+1$). We refer to those outcomes as measurement errors.
 Measurement errors 
 %(when $ZZ=-1$) 
 can be corrected in one time step as follows. First, we notice that the product of all $ZZ$ measurement operators is fixed at one, as it is the product of $ZZZZ$ stabilizers for all resource vertex states. This guarantees that the measurement errors ($ZZ=-1$) must come in pairs. As a result, one can apply (in parallel) string operators of Pauli-Xs  that connect pairs of measurement errors (see Fig.\ref{fig:1}c), bringing the post-measurement state into the ground state of the Hamiltonian defined in Eq.\ref{eq:2dtoric_post_measurement}. Physically one can regard the measurement errors as anyon excitations in the $\mathbb{Z}_2$ topological order. Applying  string operators to pair up errors can be understood as annihilating pairs of anyons, thereby giving a toric code ground state.

\subsection{More applications}	
Here we briefly discuss several classes of models that can be prepared  using our tensor network approach. We note that finite-depth adaptive protocols for preparing these states were first introduced in Ref.\cite{ashvin_2021_measurement,verresen2021_measurement_cold_atom}, albeit using a completely different approach from ours.

\textbf{Topological order away from fixed point} -- As a non-trivial application, the finite-depth local adaptive circuits can realize  non-fixed point states with topological order. This relies on the fact that a topological order can persist up to a finite  perturbation strength when the perturbation on local tensors respects  certain symmetry conditions (so that it can be mapped to local operators in the physical Hilbert space) \cite{wen_tn_symmetry_2010}. A simple example is the 2d toric code, where one considers a perturbation that breaks the $XX$ symmetry but preserves  the $ZZZZ$ symmetry in the $T$ tensor.   The $ZZZZ$ symmetry imposes the closed string condition so that the resulting  tensor-network state remains a superposition of loops. Meanwhile, the perturbation amounts to assigning a non-zero string tension, so that the wave function for each loop configuration decays exponentially with the length of loops. As shown in Ref.\cite{wen_tn_symmetry_2010}, the topological order persists up to a finite critical string tension and the transition to a trivial phase is mapped to a thermal partition function of 2d Ising model. Based on such a tensor-network construction, the non-fixed point wave function with topological order can be prepared in constant depth using the local adaptive circuits (see Appendix.\ref{appendix:non_fixed_point} for details).

\textbf{Quantum double models} -- The tensor-network-based adaptive circuits allow for a constant-depth preparation for  quantum double models of finite abelian group \cite{kitaev2003fault}. The scheme is similar to the case of 2d toric code, and one can apply a product of local unitaries to pair up measurement errors, thereby realizing the target states (see Appendix.\ref{appendix:qd} for details). Note that if one enforces the measurement outcomes, one can also prepare a quantum double of any non-abelian group. How and whether it is possible to  correct measurement outcomes efficiently for non-abelian quantum double remains an open question of our work.

\textbf{Double semion} -- An interesting application is the constant-depth preparation of a twisted $\mathbb{Z}_2$ topological order \cite{twisted_qd_2013_Wu}, i.e., the ground state of double semion model \cite{wen_string_net2005}. The state is defined on a honeycomb lattice, and is a superposition of loops weighted by a $\pm 1$ sign that depends on the number parity of loops. In Appendix.\ref{appendix:ds}, we provide an explicit preparation scheme for double semion by utilizing its exact tensor network representation \cite{wen_tnrg_2008,tn_string_net_wen_2009}.

\textbf{Fracton order} -- Finally, we consider the adaptive preparation of fracton topological phases of matter (see Ref.\cite{fracton_review_2019_hermele,you_2020_fracton} for review). These are long-range entangled quantum phases that feature fractionalized excitations with restricted mobility and a robust ground state degeneracy with system-size dependency, thereby beyond the scope of conventional topological field theories. Notably, our protocol allows for constant-depth preparation for certain fracton orders. As an illustration, in Appendix.\ref{sec:fracton} we detail the adaptive preparation for the X-cube model \cite{xcube_Chamon_2010,xcube}, a canonical type-I fracton model.

\section{Adaptive circuits via MERA}\label{sec:mera}	

\begin{figure*}[t]
    \centering
    \def\svgwidth{\textwidth}
    \begingroup%
      \makeatletter%
      \providecommand\color[2][]{%
        \errmessage{(Inkscape) Color is used for the text in Inkscape, but the package 'color.sty' is not loaded}%
        \renewcommand\color[2][]{}%
      }%
      \providecommand\transparent[1]{%
        \errmessage{(Inkscape) Transparency is used (non-zero) for the text in Inkscape, but the package 'transparent.sty' is not loaded}%
        \renewcommand\transparent[1]{}%
      }%
      \providecommand\rotatebox[2]{#2}%
      \newcommand*\fsize{\dimexpr\f@size pt\relax}%
      \newcommand*\lineheight[1]{\fontsize{\fsize}{#1\fsize}\selectfont}%
      \ifx\svgwidth\undefined%
        \setlength{\unitlength}{928.14965436bp}%
        \ifx\svgscale\undefined%
          \relax%
        \else%
          \setlength{\unitlength}{\unitlength * \real{\svgscale}}%
        \fi%
      \else%
        \setlength{\unitlength}{\svgwidth}%
      \fi%
      \global\let\svgwidth\undefined%
      \global\let\svgscale\undefined%
      \makeatother%
      \begin{picture}(1,0.34968225)%
        \lineheight{1}%
        \setlength\tabcolsep{0pt}%
        \put(0,0){\includegraphics[width=\unitlength,page=1]{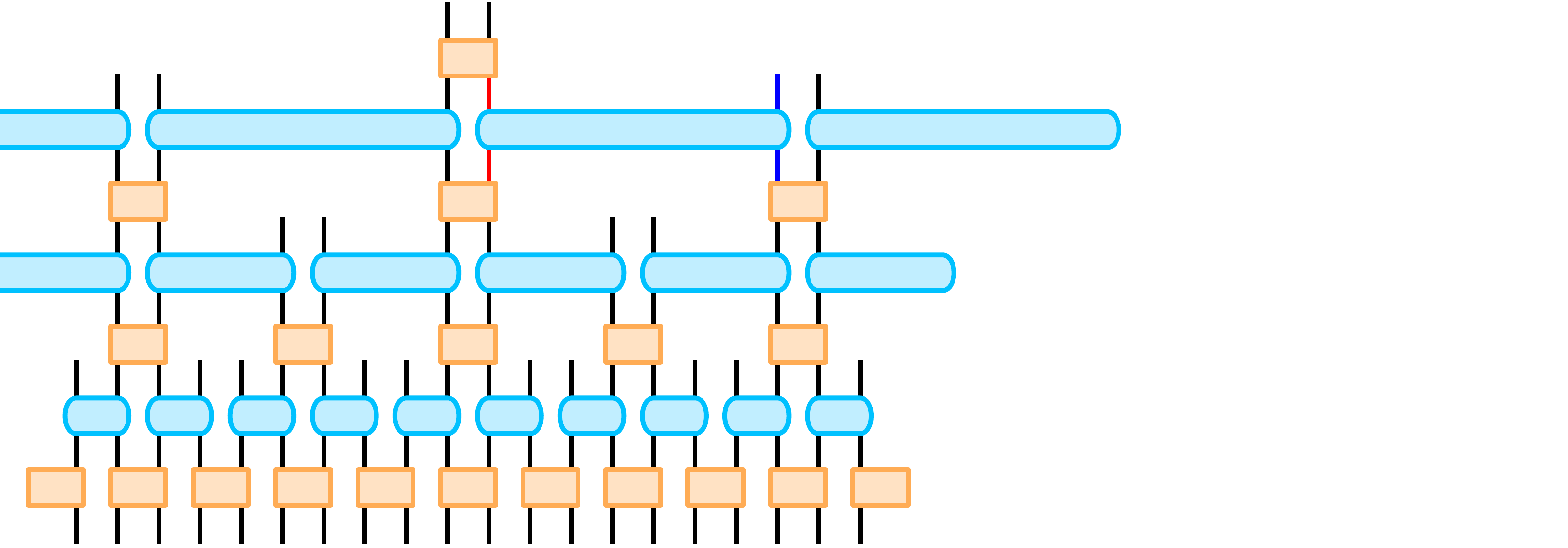}}%
        \put(0.06185146,0.0785756){\makebox(0,0)[t]{\lineheight{1.25}\smash{\begin{tabular}[t]{c}\large $w$\end{tabular}}}}%
        \put(0.08815727,0.03292915){\makebox(0,0)[t]{\lineheight{1.25}\smash{\begin{tabular}[t]{c}\large $u$\end{tabular}}}}%
        \put(0.04869993,0.12623315){\makebox(0,0)[t]{\lineheight{1.25}\smash{\begin{tabular}[t]{c}$\ket{0}$\end{tabular}}}}%
        \put(0.12761452,0.12623315){\makebox(0,0)[t]{\lineheight{1.25}\smash{\begin{tabular}[t]{c}$\ket{0}$\end{tabular}}}}%
        \put(0.15391936,0.12623315){\makebox(0,0)[t]{\lineheight{1.25}\smash{\begin{tabular}[t]{c}$\ket{0}$\end{tabular}}}}%
        \put(0.23283386,0.12623315){\makebox(0,0)[t]{\lineheight{1.25}\smash{\begin{tabular}[t]{c}$\ket{0}$\end{tabular}}}}%
        \put(0.25913869,0.12623315){\makebox(0,0)[t]{\lineheight{1.25}\smash{\begin{tabular}[t]{c}$\ket{0}$\end{tabular}}}}%
        \put(0.33805321,0.12623315){\makebox(0,0)[t]{\lineheight{1.25}\smash{\begin{tabular}[t]{c}$\ket{0}$\end{tabular}}}}%
        \put(0.36435802,0.12623315){\makebox(0,0)[t]{\lineheight{1.25}\smash{\begin{tabular}[t]{c}$\ket{0}$\end{tabular}}}}%
        \put(0.44327259,0.12623315){\makebox(0,0)[t]{\lineheight{1.25}\smash{\begin{tabular}[t]{c}$\ket{0}$\end{tabular}}}}%
        \put(0.46957741,0.12623315){\makebox(0,0)[t]{\lineheight{1.25}\smash{\begin{tabular}[t]{c}$\ket{0}$\end{tabular}}}}%
        \put(0.5484919,0.12623315){\makebox(0,0)[t]{\lineheight{1.25}\smash{\begin{tabular}[t]{c}$\ket{0}$\end{tabular}}}}%
        \put(0.18022419,0.21752608){\makebox(0,0)[t]{\lineheight{1.25}\smash{\begin{tabular}[t]{c}$\ket{0}$\end{tabular}}}}%
        \put(0.20652895,0.2175262){\makebox(0,0)[t]{\lineheight{1.25}\smash{\begin{tabular}[t]{c}$\ket{0}$\end{tabular}}}}%
        \put(0.39066286,0.2175262){\makebox(0,0)[t]{\lineheight{1.25}\smash{\begin{tabular}[t]{c}$\ket{0}$\end{tabular}}}}%
        \put(0.41696765,0.2175262){\makebox(0,0)[t]{\lineheight{1.25}\smash{\begin{tabular}[t]{c}$\ket{0}$\end{tabular}}}}%
        \put(0.49588219,0.30881916){\makebox(0,0)[t]{\lineheight{1.25}\smash{\begin{tabular}[t]{c}$\ket{0}$\end{tabular}}}}%
        \put(0.52218699,0.30881916){\makebox(0,0)[t]{\lineheight{1.25}\smash{\begin{tabular}[t]{c}$\ket{0}$\end{tabular}}}}%
        \put(0.07500481,0.30881916){\makebox(0,0)[t]{\lineheight{1.25}\smash{\begin{tabular}[t]{c}$\ket{0}$\end{tabular}}}}%
        \put(0.1013096,0.30881916){\makebox(0,0)[t]{\lineheight{1.25}\smash{\begin{tabular}[t]{c}$\ket{0}$\end{tabular}}}}%
        \put(0.40381439,0.26274349){\makebox(0,0)[t]{\lineheight{1.25}\smash{\begin{tabular}[t]{c}Right panel\end{tabular}}}}%
        \put(0,0){\includegraphics[width=\unitlength,page=2]{MERAFigures.pdf}}%
        \put(0.94023484,0.08963358){\makebox(0,0)[t]{\lineheight{1.25}\smash{\begin{tabular}[t]{c}$\scriptstyle  a'_3 b'_3$\end{tabular}}}}%
        \put(0,0){\includegraphics[width=\unitlength,page=3]{MERAFigures.pdf}}%
        \put(0.72753819,0.23555948){\makebox(0,0)[t]{\lineheight{1.25}\smash{\begin{tabular}[t]{c}$\scriptstyle  a_1 b_1$\end{tabular}}}}%
        \put(0.81267957,0.23587667){\makebox(0,0)[t]{\lineheight{1.25}\smash{\begin{tabular}[t]{c}$\scriptstyle  a_2 b_2$\end{tabular}}}}%
        \put(0.89784092,0.23548921){\makebox(0,0)[t]{\lineheight{1.25}\smash{\begin{tabular}[t]{c}$\scriptstyle  a_3 b_3$\end{tabular}}}}%
        \put(0.68494351,0.08970388){\makebox(0,0)[t]{\lineheight{1.25}\smash{\begin{tabular}[t]{c}$\scriptstyle  a'_0 b'_0$\end{tabular}}}}%
        \put(0.77008489,0.09002107){\makebox(0,0)[t]{\lineheight{1.25}\smash{\begin{tabular}[t]{c}$\scriptstyle  a'_1 b'_1$\end{tabular}}}}%
        \put(0.85524622,0.08963358){\makebox(0,0)[t]{\lineheight{1.25}\smash{\begin{tabular}[t]{c}$\scriptstyle  a'_2 b'_2$\end{tabular}}}}%
        \put(0,0){\includegraphics[width=\unitlength,page=4]{MERAFigures.pdf}}%
        \put(0.63819143,0.05425283){\makebox(0,0)[t]{\lineheight{1.25}\smash{\begin{tabular}[t]{c}$v'$\end{tabular}}}}%
        \put(0,0){\includegraphics[width=\unitlength,page=5]{MERAFigures.pdf}}%
        \put(0.96141965,0.22591824){\makebox(0,0)[t]{\lineheight{1.25}\smash{\begin{tabular}[t]{c}$v$\end{tabular}}}}%
        \put(0,0){\includegraphics[width=\unitlength,page=6]{MERAFigures.pdf}}%
        \put(0.60590499,0.31945962){\makebox(0,0)[t]{\lineheight{1.25}\smash{\begin{tabular}[t]{c}\large($b$)\end{tabular}}}}%
        \put(0.03055145,0.31945962){\makebox(0,0)[t]{\lineheight{1.25}\smash{\begin{tabular}[t]{c}\large($a$)\end{tabular}}}}%
        \put(0,0){\includegraphics[width=\unitlength,page=7]{MERAFigures.pdf}}%
        \put(0.59610281,0.13382665){\makebox(0,0)[t]{\lineheight{1.25}\smash{\begin{tabular}[t]{c}Time\end{tabular}}}}%
        \put(0.66604031,0.3300666){\makebox(0,0)[t]{\lineheight{1.25}\smash{\begin{tabular}[t]{c}\large$A'$\end{tabular}}}}%
        \put(0.96364012,0.3300666){\makebox(0,0)[t]{\lineheight{1.25}\smash{\begin{tabular}[t]{c}\large$B'$\end{tabular}}}}%
        \put(0.6362651,0.30365821){\makebox(0,0)[t]{\lineheight{1.25}\smash{\begin{tabular}[t]{c}\large$A$\end{tabular}}}}%
        \put(0.93386496,0.30365823){\makebox(0,0)[t]{\lineheight{1.25}\smash{\begin{tabular}[t]{c}\large$B$\end{tabular}}}}%
        \put(0.94502912,0.15169624){\makebox(0,0)[t]{\lineheight{1.25}\smash{\begin{tabular}[t]{c}\large \slantbox{$w$}\end{tabular}}}}%
        \put(0.64554946,0.23426831){\makebox(0,0)[t]{\lineheight{1.25}\smash{\begin{tabular}[t]{c}\slantbox{$\scriptstyle a_0 b_0$}\end{tabular}}}}%
      \end{picture}%
    \endgroup%
    \caption{MERA state preparation using local adaptive circuits. (a) a MERA circuit for state preparation with isometries $w$ and disentanglers $u$ starting from the product state $\otimes_i \ket{0}_i$. Time goes from top to bottom and the qubits are arranged horizontally on a 1d lattice. While we present a particular architecture of MERA where only the isometries $w$ are non-local, our adaptive circuits can be straightforwardly generalized to other MERA where both  isometries and unitaries are non-local. (b) Local adaptive circuit for implementing the non-local isometry indicated in the left panel. The ancilla chain starts with a series of Bell pairs (wiggly lines). They serve as a  resource to teleport the state at site $A$ (in red) to the ancilla site $B'$, which is next to  site $B$ (in blue). After applying the  local isometry gate between $B$ and $B'$, we teleport the state in $B'$ back to $A$, and the Bell pairs return to their original positions.  All isometries in a given layer of MERA can be implemented in this way in parallel.}
    \label{fig:MERAFigures}
\end{figure*}

Our second class of adaptive circuits realizes a multiscale entanglement renormalization ansatz (MERA) \cite{vidal_2007_mera,vidal_2008_mera} using (spatially) \textit{local} unitary gates and local Bell-pair measurements. In contrast, a naive implementation of MERA circuits requires \textit{non-local} unitary gates to generate entanglement at various length scales. Since a MERA can characterize both gapped topological orders and critical states described by a conformal field theory (CFT), both of these classes of states can be prepared in $O(\log L)$-depth local adaptive circuits with $L$ being the system size. In particular, the $O(\log L)$-depth adaptive MERA circuit is optimal for preparing a 1d critical state with $O(\log L)$ entanglement scaling (see Appendix.\ref{appendix:entanglement_bound} for proof).

MERA is a framework for real-space renormalization group (RG) transformations, which systematically remove short-range entanglement at various length scales, thereby extracting the long-distance, universal structure of correlations and entanglement in quantum systems. MERA consists of alternating layers of unitary gates (dubbed disentanglers) and isometries. At each scale, unitary gates $u$ are applied to remove the short-range entanglement, followed by isometries $w$ that perform coarse graining. A schematic of MERA is presented in Fig.\ref{fig:MERAFigures}a, where the RG flows from the bottom (UV) to the top (IR). On the other hand, by reading the structure from top to bottom, one obtains a circuit that prepares a target state. In particular, the isometry $w$ can be regarded as a unitary that takes as input two spins, one of which is in the state $\ket{0}$. However, such a circuit requires spatially non-local gates, which obstructs its implementation. Our adaptive protocol overcomes this difficulty with the aid of Bell measurement.

The main insight of our construction relies on quantum teleportation \cite{wootters_1993_teleport}. In the simplest setting, one considers three qubits labeled by $A,B,C$. Given an unknown state $\ket{\psi}= m_0 \ket{ 0} +  m_1 \ket{ 1} $ on $A$, one would like to accomplish the task of teleportation, after which C is in the state $\ket{\psi}$. This is achieved by first preparing a Bell pair $\ket{\Psi_{00}}_{BC} = \frac{1}{\sqrt{2}}( \ket{ 00}_{BC} + \ket{11}_{BC})$, followed by a Bell pair measurement on $A,B$. This projects $AB$ to one of the four possible Bell pairs: $\ket{\Psi_{\alpha \beta}}_{AB} =( \mathbb{I}\otimes Z^{\alpha } X^{\beta})\ket{\Psi_{00}}_{AB}$ with $\alpha,\beta \in \{0,1\}$, and meanwhile, the state in $C$ becomes $Z^{\alpha}X^{\beta} \ket{\psi}_C$. Finally, applying the unitary correction $(Z^{\alpha}X^{\beta})^{\dagger}$ on $C$ achieves the desired teleportation.

Employing the above idea, we devise a local adaptive protocol for implementing non-local gates, which are essential for realizing MERA circuits. The protocol in one space dimension is summarized as follows (see also Fig.\ref{fig:MERAFigures}b). We consider a 1d lattice of qubits (physical chain), next to which we append a 1d lattice of ancilla qubits (ancilla chain) that form adjacent Bell pairs. To apply a unitary gate on two distant physical qubits $A$ and $B$, we perform simultaneous non-overlapping Bell measurements to teleport the information encoded in the physical qubit $A$ through the ancilla chain to an ancilla qubit $B'$ next to $B$ in constant time. We apply a desired unitary gate acting on $B$ and $B'$, and then we teleport the state encoded in $B'$ back to the physical qubit $A$. With this, a non-local unitary gate acting on two qubits $A$ and $B$ can be implemented in constant time independent of their separation. Note that the procedure above can be straightforwardly generalized to higher dimensions: given a physical system on a $d$-dim lattice, one can append a $d$-dim ancilla lattice consisting of Bell pairs. The ancilla lattice serves as a portal for  teleporting a physical qubit to any qubit in the ancilla lattice in constant time, and therefore, any spatially non-local gate can be implemented analogously.

\subsection{Protocol}\label{sec:mera_protocol}
Now we specialize in the MERA circuits depicted in Fig. \ref{fig:MERAFigures} for preparing states of qubits arranged in a 1d lattice. Note that while the depicted MERA presents a structure where only isometries $w$ are non-local, our adaptive circuits can be easily adapted to other MERA where both unitaries $u$ and isometries $w$ are non-local. 

Since any unitary $u$ is local, below we focus on the implementation of non-local isometries $w$. To start, we introduce an ancilla qubit for every qubit in the original system, yielding an ancilla chain attached to the original chain.

We would like to perform an isometry $w$ acting on two distant physical qubits - one labeled by $A$ at lattice coordinates $x=0$ and one labeled by $B$ at lattice coordinates $x=2n+1$ with $n \in \mathbb{Z}$. For instance, the qubits $A$ and $B$ are colored in red and blue with coordinates $0$ and $2n+1$ for $n=3$ respectively in Fig.\ref{fig:MERAFigures}b. We will teleport the state of the physical qubit $A$ to the ancilla qubit $B'$ next to the physical qubit $B$, i.e. $B'$ has the same coordinate $2n+1$ but in the ancilla chain. To this end, one initializes the ancilla qubits of coordinates $x=0,1,\ldots, 2n+1$ in the tensor product of adjacent Bell pairs $\otimes_{i=0}^{n} \ket{\Psi_{\alpha_i\beta_i}}_{2i,2i+1}$, where $\ket{\Psi_{\alpha \beta}} = \frac{1}{\sqrt{2}} (1 \otimes v_{\alpha \beta}) (\ket{00} + \ket{11})$ with $v_{\alpha \beta} = Z^\alpha X^\beta$ and $\alpha,\beta\in \{0,1 \}$. We then  simultaneously perform non-overlapping Bell measurements acting on (i) a pair of physical qubit $A$ and ancilla qubit $A'$ at coordinate $x=0$, and (ii) the pairs $(2i-1,2i)$ for $i=1,2,\ldots, n$ in the ancilla chain. It follows that the qubits $A, A'$ are projected to a Bell pair $\ket{\Psi_{a_0b_0}}_{AA'}$, and the qubit pairs $(2i-1,2i)$ are projected to $\ket{\Psi_{a_i b_i}}_{2i-1,2i}$ for $i=1,2,\ldots, n$. At the same time, the state at site $A$ is  teleported to site $B'$ up to a unitary correction $v = \left(\prod_{i=0}^n v_{a_i b_i} v_{\alpha_i \beta_i} \right)^\dagger$.

After the above teleportation step, one applies the local isometry gate $w$ on $B'$ and $B$. Finally, we simultaneously measure the pairs of qubits $(2i,2i+1)$ for $i=0,\cdots,n$ in the ancilla chain. The measurement projects those qubits to Bell pairs $\otimes_{i=0}^{n} \ket{\Psi_{a'_i b'_i}}_{2i,2i+1}$, and the state in $B'$ is teleported back to $A$ up to a unitary correction determined by the outcomes $(a'_i,b'_i)$ in a similar way to the previous teleportation. The procedure above, therefore, realizes a non-local isometry gate in constant time with the ancilla chain as a teleportation route\footnote{While we only detail the implementation of a non-local gate acting on site $0$ and site $2n+1$, we note that a non-local gate on any two sites $A$ and $B$ can be implemented analogously since the ancilla spin chain provides a portal to teleport the physical qubit $A$ to an ancilla qubit $B'$ that is $O(1)$ close to $B$.}. Importantly, for each layer consisting of  non-local gates, the above teleportation procedures occur in different  non-overlapping regions of the ancilla chain. This means that the aforementioned protocol for a single isometry gate can be applied in parallel to all isometry gates at the same time. Therefore, every layer of a MERA circuit is implemented with local unitaries and local measurements in constant time.  Below we discuss two non-trivial applications of our adaptive circuit.

\subsection{Critical spin chain}
MERA provides a natural description for gapless spin chains characterized by 1+1D CFTs. While obtaining a MERA in general requires variational optimization over  the isometries $w$ and unitaries $u$, it was found that using the wavelet theory, $w$ and $u$ can be constructed  analytically for spin models that can be mapped to free-fermions. A notable example is the ground state of a  critical quantum Ising chain, where a MERA of bond dimension 2 already well approximates the actual ground state and the scaling dimension of certain operators can even be exactly reproduced \cite{white_2016_wavelet}. 

Furthermore, expectation values of local observables can be approximated with an error that decays exponentially with the number of layers within a fixed length scale \cite{Haegeman2018}. While the exact form of the quantum circuit in Fig.~\ref{fig:MERAFigures} is different from the one introduced in Ref.~\cite{Haegeman2018}, the two are related to each other up to a local swapping of qubits. Since this is also a two-qubit gate, the circuit in Ref.~\cite{Haegeman2018} can be reproduced by the circuit in Fig.~\ref{fig:MERAFigures} up to at most a two-fold increase in the overall depth of the circuit.

\subsection{Gapped topological order}
While our protocol in Sec.\ref{sec:mera_protocol} focuses on implementing a MERA circuit in 1d, one can straightforwardly adopt the same idea to implement a 2d MERA circuit by appending a 2d lattice of ancilla qubits in Bell pairs as a portal for teleporting physical qubits. This enables  the preparation of quantum doubles of any finite group $G$ in $O(\log L)$ depth via their exact MERA construction \cite{vidal_2008_mera_to}. Remarkably, while previously known low-depth preparations for quantum doubles in adaptive circuits are limited to solvable non-abelian groups \cite{ashvin_2021_measurement,verresen2021_measurement_cold_atom,bravyi_2022_adaptive}, the group $G$ in our protocol can be any finite non-abelian group. More generally, this class of adaptive circuits can prepare in log depth any Levin-Wen string-net model \cite{wen_string_net2005} based on its exact MERA construction \cite{vidal_2009_mera_string_net}. In these cases, the local Hilbert space corresponds to a $d$-dimensional qudit. By employing the measurement-based teleportation scheme for qudits \cite{wootters_1993_teleport}, one can implement any non-local unitary gate using local unitary gates and measurements, thereby allowing us to prepare any quantum double or string-net state in $O(\log L)$-depth.

\section{Adaptive circuits via partons}\label{sec:partons}
Here we introduce a class of adaptive circuits inspired by parton constructions, a widely-used  approach to  characterize the emergence of topological order in various spin systems. Notably, this allows us to efficiently prepare a chiral non-abelian topological order. We will introduce this class of circuits in the context of the Kitaev honeycomb model \cite{kitaev_2006}, a paradigmatic example that is exactly solvable via parton construction. Consider a honeycomb lattice with qubits on sites and the Hamiltonian 
\begin{equation}\label{eq:honeycomb}
H=   - J_x \sum_{ \text{x-links}} X_j X_k  - J_y\sum_{ \text{y-links}} Y_jY_k  - J_z\sum_{\text{z-links}} Z_j Z_k, 
\end{equation}
where $\alpha$-links with $\alpha=x, y, z $ label links of three different orientations. This model can be solved by fractionalizing each spin on site $j$ into four Majoranas $b^x_j,b^y_j,b^z_j,c_j$: $X_j = i b^x_j c_j $, $Y_j = i b^y_j c_j $, $Z_j = i b^z_j c_j $ subject to the constraint $D_j = b_j^x b_j^yb_j^z  c_j   =1 $ so that the spin operators satisfy  $X_j Y_j Z_j = i $. In terms of  these Majorana partons, the Hamiltonian can be written as 

\begin{figure}
	\centering
	\begin{subfigure}{0.47\textwidth}
		\includegraphics[width=\textwidth]{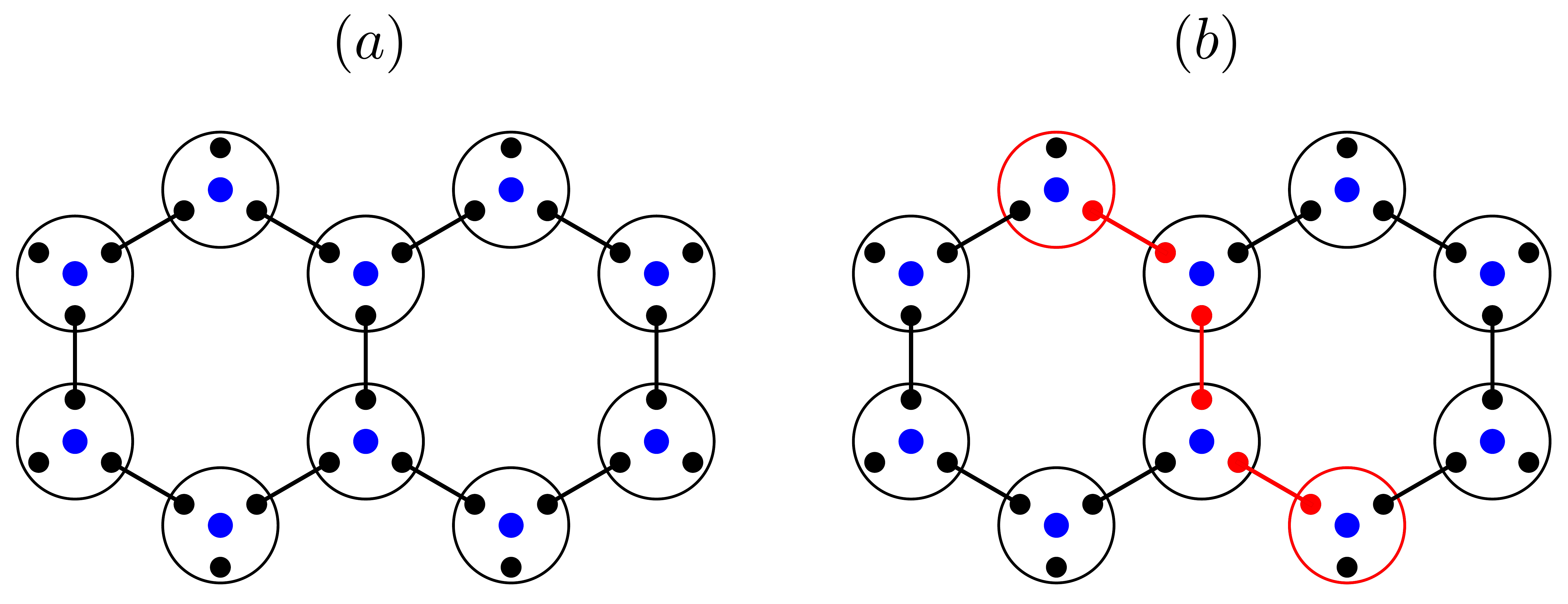}
	\end{subfigure}
	\caption{Adaptive circuits via parton construction. (a) Given the Kitaev honeycomb model (Eq.\ref{eq:honeycomb}) with qubits defined on sites, each qubit is fractionalized into four Majorana fermions subject to local constraints $D_j = b_j^x b_j^y b_j^z c_j=1$.  The Hamiltonian is mapped to a model of quadratic $c$ Majoranas coupled to a background $\mathbb{Z}_2$ gauge field represented by $b$ Majorana dimers. The physical ground state is obtained by projecting the free-fermion solution into the physical subspace with $D_j=1$. This motivates the adaptive circuit: preparing a free-fermion ground state, one simultaneously measures $D_j$ operators. The desired ground state of the Kitaev model is obtained when every outcome is $D_j=1$. (b) Measurement errors ($D_j=-1$) appear in pairs and can be corrected by applying a string operator (the product of $b$ Majoranas) that connects two errors. Any pattern of errors is corrected in a single time step by simultaneously applying string operators.}
	\label{fig:parton}  
\end{figure}

\begin{equation}
H=  \frac{i}{4} \sum_{\expval{jk}}  (2J_{jk} u_{jk} )c_j c_k,
\end{equation}
where $J_{jk}= J_{\alpha}$ and $u_{jk}= i b^{\alpha}_{j}b_k^{\alpha}$ with $\alpha \in \{x,y,z\}$ depending on the orientation of the link $\expval{jk}$. Importantly, the operators $u_{jk}$ commute with each other and the Hamiltonian $H$, and therefore, the Hilbert space splits into the sectors organized by the eigenvalues $u_{jk}=\pm 1$. For each sector, the Hamiltonian is quadratic in $c$ Majoranas, whose ground state can be exactly determined. The sector with the lowest energy is the flux-free subspace where the product of $u_{jk}$ around each plaquette is one. Therefore, one may choose $u_{jk}=1$ on all links and find the ground state $\ket{\psi_0}$ of the resulting quadratic $c$-Majorana Hamiltonian.  Finally, $\ket{\psi_0}$ is projected to the physical subspace specified by $D_j=1$ to obtain the physical ground state of spin Hamiltonian

\begin{equation}\label{eq:parton}
\ket{\psi} =  \prod_j \frac{1+D_j }{2} \ket{\psi_0}.
\end{equation}

Using the parton construction, Ref.\cite{kitaev_2006} found two distinct phases depending on the choice of $J_x, J_y, J_z$: gapped phase (A phase) exhibiting a $\mathbb{Z}_2$ topological order and gapless phase (B phase). Importantly, applying a magnetic field, i.e. a time-reversal symmetry-breaking perturbation, opens a gap in the B phase and leads to a chiral topological order that hosts non-abelian Ising anyons. Specifically, the effective parton Hamiltonian derived from perturbation theory reads 
\begin{equation}\label{eq:c_H}
H_{\text{eff}} =  \frac{i}{4} \sum_{jk} A_{jk} c_{j}c_k
\end{equation}
with $A_{jk}$ including both  nearest-neighbor and next-nearest-neighbor hopping amplitudes (see Ref.\cite{kitaev_2006} for details). The band structure of $c$ Majorana fermions is described by a $p_x\pm ip_y$ superconductor ($\pm 1$ sign depends on the direction of the applied magnetic field), which supports gapped fermionic excitations, and crucially, binds an unpaired Majorana mode to each vortex \cite{Read_pwave_2000,pwave_2001_Ivanov}. As a result, by projecting to the physical subspace, the ground state exhibits a gapped, chiral topological order characterized by an Ising anyon theory due to the gapped fermionic excitations and gapped vortex excitations with non-abelian (Majorana) statistics.

The above parton construction immediately suggests a protocol for preparing the Ising-anyon topological order: starting from a free-fermion state consisting of neighboring b-Majorana dimers and the ground state of the c-Majorana hopping Hamiltonian (Eq.\ref{eq:c_H}), one simultaneously measures the operator $D_j= b^x_jb^y_jb^z_jc_j$ on every vertex. With the measurement outcomes $D_j=1$ on every vertex, one exactly obtains the state in the form of Eq.\ref{eq:parton}, thereby exhibiting a non-abelian Ising anyon topological order (see Fig.\ref{fig:parton}a). One caveat is the following: strictly-local unitary circuits of finite depth cannot prepare c-Majoranas in a p+ip superconducting state due to its non-zero Chern number. On the other hand, being a gapped invertible phase, it can be prepared in finite time via adiabatic evolution. Specifically, given a gapped Hamiltonian $H_A$ whose gapped ground state is a p+ip superconductor, one can find a gapped Hamiltonian $H_B$ whose ground state is a p-ip superconductor. Since they are the inverse of each other, one can adiabatically turn on the interaction $H_{AB}$ to obtain a trivial state without closing the gap in finite time. This in turn implies that starting from a trivial state, one can adiabatically prepare decoupled p+ip and p-ip superconductors in finite time, and then one can take one of them as a resource to implement our aforementioned protocol.

Now we discuss how to correct the measurement outcome from $D_j=-1$ to $D_j=1$. We first notice that the product of all measurement operators, i.e. $\prod_jD_j$, is a conserved quantity solely fixed by the state before measurements  (i.e.$\ket{\psi_0}$). This is because by reshuffling the $b$ Majoranas and expressing  them in the conserved quantity $u_{ij}$, one finds $\prod_j D_j  \sim \prod_{\expval{ij}}u_{ij} \prod_{i}c_i$, where $\sim$ denotes an equal sign up to a phase, which has been exactly derived in Ref.\cite{loss_2011_honeycomb}. Since $\prod_i c_i$ simply measures the c fermion parity of $\ket{\psi_0}$, $\prod_j D_j$ is fixed by the initial state $\ket{\psi_0}$. Thus, given the initial state $\ket{\psi_0}$  with $\prod_{j}D_j=1$, measurement errors ($D_j =-1$) arise in pairs, and one can apply a string operator consisting of b Majoranas to correct a pair of errors (see Fig.\ref{fig:parton}b). As a result, any measurement errors can be corrected in one step to obtain the desired target state. Note that this is consistent with the fact that the gauge field represented by $b$ Majoranas is abelian (i.e. $\mathbb{Z}_2$) despite the existence of non-abelian topological order. 

We note that a low-depth preparation of chiral Ising topological order  via measurements has been proposed in Ref.\cite{ashvin_2021_measurement}, which nevertheless differs from our protocol. The protocol in Ref.\cite{ashvin_2021_measurement} involves (1) preparing the fermions in a p+ip superconductor (2) adding bosonic spin degrees of freedom followed by a depth-1 unitary circuit that couples bosonic spins and fermions and (3) measuring fermion parity operators, and applying a depth-1 unitary to correct measurement outcomes. Finally, one obtains the chiral Ising topological order for bosonic spins. Conceptually, it can be understood as gauging the fermion parity (equivalently a higher-dimensional Jordan-Wigner transformation or bosonization) for the p+ip superconductor. 

In contrast, our protocol only takes (Majorana) fermions as input without involving bosonic spins. This differs from the perspective of gauging fermion parity, in which case bosonic degrees of freedom are needed to minimally couple to fermions. However, measuring fermions projects the fermionic Hilbert space to a subspace which does allow for a bosonic description.  Our approach is more akin to the ``topological bootstrap'' \cite{topbootstrap}, involving Kondo coupling between a parton free fermion state and a spin system, which can realize the same topological order as the projected parton state in the spin system.  Here, instead of Kondo coupling, the projective measurements play the role of projecting onto an effective spin Hilbert space.

\begin{figure*}[t]
	\centering
	\begin{subfigure}{1.0\textwidth}
		\includegraphics[width=\textwidth]{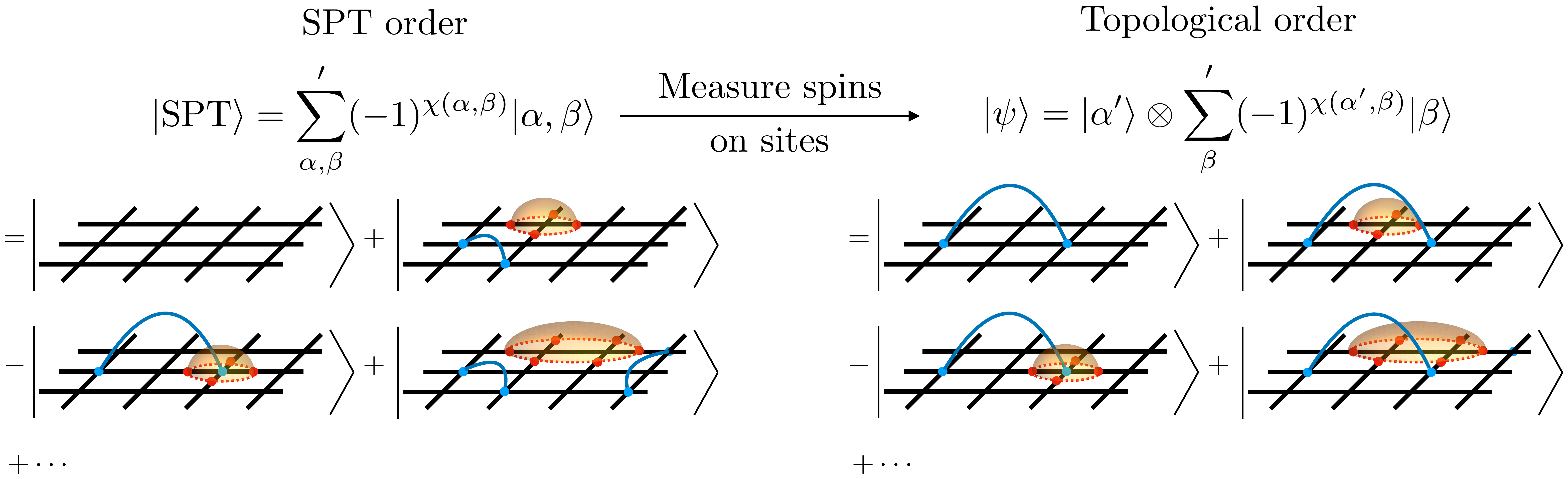}
	\end{subfigure}
	\caption{Adaptive circuits from  disentangling perspective -- The schematic shows a 2d (SRE) cluster state with a $\mathbb{Z}_2\times\mathbb{Z}_2$ SPT order, which is understood as two species of fluctuating domain walls with non-trivial braiding sign structure in the symmetry-charge (Pauli-X) basis. The sites on the boundary of blue open strings form a domain wall of a classical Ising gauge theory, and the loops (defined in dual lattice) on the boundary of red open membranes form a domain wall of a classical Ising theory. The wave functions take the value 1 (-1) corresponding to the even (odd) number  of times that blue strings pierce through red membranes. Measuring the spins on sites projects those spins into a specific product state. The unmeasured spins form a superposition of loops in the dual lattice up to a $\pm 1$ sign, which can be further removed by applying string operators, leading to a 2d toric code state with $\mathbb{Z}_2$ topological order.} 
	\label{fig:disentangle}  
\end{figure*}

\section{Adaptive circuits via disentangling fluctuating domain walls}\label{sec:disentangling}
Finally, we build on a class of local adaptive circuits in the recent works \cite{ashvin_2021_measurement,verresen2021_measurement_cold_atom}, which point out the emergence of long-range order by measuring certain SRE states with  SPT order.  We will discuss such correspondence using an explicit wave-function picture with a ``disentangling'' perspective. The central insight is as follows:  wavefunctions of certain SRE states  can be described by two species of fluctuating domain walls with a non-trivial braiding phase. As a result, measuring one species can disentangle the braiding between two species: the measured species  is projected to a trivial product state while the un-measured species remains a wave function of fluctuating domain walls, i.e. a superposition of all domain-wall configurations, thereby exhibiting certain long-range order and entanglement.

The above idea leads to a simple understanding of the emergence of  a $\mathbb{Z}_2$ LRE topological order by   measuring an SRE state with SPT order protected by a $\mathbb{Z}_2\times  \mathbb{Z}_2$ symmetry. While this result has been  recently pointed out in Ref.\cite{ashvin_2021_measurement,verresen2021_measurement_cold_atom} (see also Ref.\cite{erez_2022_duality}), our wavefunction perspective provides an intuitive understanding of the connection between SPT order and topological order via measurements. In addition, our idea immediately shows how to prepare certain symmetry-enriched topological (SET) orders by measuring SRE states. Finally, extending  the idea to domain walls without spatial locality within the framework of the chain complex enables the preparation of any CSS code in constant depth, reproducing a result in Ref.\cite{stace_2016_css}.

\subsection{From SPT to topological order}

\textbf{2d SPT with $\mathbb{Z}_2 \text{ 0-form} \times \mathbb{Z}_2 \text{ 1-form}$ symmetry} -- Here we consider a 2d cluster state to demonstrate our disentangling approach (see also the application to the 1d SPT cluster state in Appendix.\ref{sec:1d_spt}). While the emergence of long-range order via measurement in this model has been well known \cite{Raussendorf_2001_ghz}, our perspective provides a novel insight that allows for straightforward generalization as we will discuss later.

To start, we consider a 2d square lattice with every site and link accommodating a spin. The model Hamiltonian is defined as

\begin{equation}\label{eq:2d_cluster_spt}
	H_{\text{SPT}} = - \sum_{v}  X_v \prod_{ l \vert v\in \partial l     }Z_l  -  \sum_{ l  }  X_l \prod_{ v |  v \in \partial l   }Z_v.    	
\end{equation}
The first term is a product of a Pauli-X  on the site $v$ and four Pauli-Zs on links emanating from the site, and the second term is the product of a Pauli-X on the link $l$ and two Pauli-Zs on two sites located on the boundary of the link. Assuming periodic boundary conditions, $H_{\text{SPT}}$ can be obtained from a trivial paramagnet $-\sum_v X_v - \sum_l X_l$ by a finite-depth local unitary $U=\prod_{\expval{v,l}} \text{CZ}_{v,l}$, which is a product of control-Z gates $\text{CZ}$ acting on any two neighboring qubits (one on site and one on link). Correspondingly, applying $U$ to a trivial paramagnet state $\otimes_{v}\ket{+}_v \otimes_{l} \ket{+}_l$ with $\ket{+}$ labeling a $+1$ single-qubit Pauli-X eigenstate leads to the ground state of $H_{\text{SPT}}$:

\begin{equation}
\ket{\text{SPT}}  = \prod_{\expval{v,l}} \text{CZ}_{v,l} \otimes_{v}\ket{+}_v \otimes_{l} \ket{+}_l 
\end{equation}

$\ket{\text{SPT}} $ exhibits an SPT order protected by a $\mathbb{Z}_2\times \mathbb{Z}_2$ symmetry, where the first $\mathbb{Z}_2$ is a 0-form symmetry acting on all sites, i.e. $\prod_vX_v$, and the second  $\mathbb{Z}_2$ is a 1-form symmetry acting on links that form deformable 1d closed loops $\mathcal{C}$, i.e. $\prod_{ l \in \mathcal{C}}X_l$.  $\ket{\text{SPT}}$ is a non-trivial SPT as it cannot be transformed to a trivial product state using any finite-depth unitary local circuit that respects the symmetry.

Our perspective begins by expressing the  SPT  in the Pauli-X basis, i.e. the symmetry-charge basis in which symmetry action is diagonal \cite{zohar_2015_spt,zohar_2016_spt}: $\ket{\alpha, \beta} \equiv \ket{   \{\alpha_v\}, \{\beta_l\}      }$ with $ \alpha_v = \pm 1$ corresponding to the $\pm 1 $ eigenvalue of $X_v$ acting on sites (similarly for $\beta_l$ defined on links). We find $\ket{\text{SPT}}$ in $X$ basis can be written as (up to a normalization)

\begin{equation}
	\ket{\text{SPT}}  =   \sum'_{\alpha, \beta  }   (-1)^{ \chi(\alpha, \beta )  }   \ket{  \alpha , \beta   }.
\end{equation}
Here $\sum'$ indicates that only those product state bases fulfilling certain constraints are allowed. Specifically, the $\mathbb{Z}_2$ 0-form symmetry $\prod_vX_v=1$ imposes the constraint that the allowed $\alpha$ configurations satisfy $\prod_v \alpha_v=1$, and therefore, the spins with $\alpha_v=-1$ must come in pairs. It is useful to regard $\alpha$  as a domain wall of the 2d classical $\mathbb{Z}_2$ gauge theory 
% $- \sum_v \prod_{l , v\in \partial l} = - \sum_{v}ZZZZ$ 
with four Pauli-Zs on links emanating from a vertex as a local term. A domain wall configuration $\alpha $ then corresponds to a violation of local terms, which must come in pairs on the boundary of open strings.  On the other hand, the $\mathbb{Z}_2$ 1-form symmetry imposes the constraint  $\prod_{l \in \mathcal{C}} \beta_l=1$, implying that $\beta$ must be a configuration of closed loops in the ``dual'' lattice. One can regard $\beta$ as a  domain wall of the 2d Ising model
% $-\sum_{l} \prod_{v| v\in \partial l}Z = -\sum_{l}ZZ$
with Pauli-Z interactions between neighboring spins on sites.  Each domain wall is a closed loop in the dual lattice on the boundary of an open membrane. Importantly, $\chi(\alpha, \beta)$ is the linking number between two species of domain wall configurations, i.e. it counts the number of times that  a string pierces through the membrane, see Fig. \ref{fig:disentangle}.

Given  the SPT state above, performing Pauli-X measurements on sites gives the following state 

\begin{equation}\label{eq:2dspt_post_measure}
\ket{\psi}  =  \ket{ \alpha' } \otimes  \sum'_{\beta}   (-1)^{ \chi(\alpha', \beta )  }   \ket{   \beta },
\end{equation}
where the measured species is randomly projected to a particular $\alpha'$ configuration, and crucially, the un-measured species $\beta$ forms a superposition of closed loops in the dual lattice up to signs $(-1)^{\chi(\alpha', \beta)}$. To remove the signs, one needs to find an operator $O$ acting on links such that  $O\ket{\beta} = (-1)^{ \chi (\alpha',\beta)} \ket{\beta}$. This is achieved by choosing $O$ to be a product of $X$ operators on open strings whose boundary is the domain wall $\alpha'$ of the 2d $\mathbb{Z}_2$ gauge theory. After this sign-removal procedure, one then obtains  the exact $\mathbb{Z}_2$ toric code ground state for spins on links, i.e. $\sum_{\beta}' \ket{ \beta}$. Note that the post-measurement state exhibits a $\mathbb{Z}_2$ topological order even without removing the sign, since the correcting unitary is a finite depth circuit. Without correction, the state can be regarded as an excited state of the toric code, and the anyon excitations live on those vertices with $X_v=-1$ in the $\alpha'$ configuration.

It is also interesting to measure Pauli-Xs on links of $\ket{\text{SPT}}$. This projects those spins to a particular Ising domain-wall $\beta'$. By further applying membrane operators of Pauli-Xs  acting on sites whose boundary is $\beta'$, the spins on sites exactly form a many-body cat (GHZ) state $\sum'_{\alpha}  \ket{\alpha} \sim  (1+ \prod_v X_v) \sum_{\alpha} \ket{\alpha}  = \ket{\uparrow \uparrow \uparrow \cdots}  +  \ket{ \downarrow\downarrow\downarrow\cdots}$, which exhibits a (spontaneous symmetry breaking) long-range order, i.e. $\expval{Z_{v} Z_{v'} }=1$ for any two spins.

\textbf{3d SPT with $\mathbb{Z}_2 \times \mathbb{Z}_2$  1-form symmetry} -- Another illuminating  example is the 3d (SPT) cluster state \cite{3d_cluster_state_2005}.  Consider a 3d cubic lattice where each link and each plaquette (face) accommodates a qubit, with periodic boundary conditions in all space directions, the model Hamiltonian is $H_{\text{SPT}}=- \sum_p X_p \prod_{ l \in \partial p} Z_l  - \sum_l X_l \prod_{p, \partial p \ni l} Z_l$. The first term is a product of Pauli-X on a plaquette and four Pauli-Zs on links around the boundary of the plaquette, and the second term is a product of Pauli-X on a link and four Pauli-Zs on plaquettes adjacent to the link. $H_{\text{SPT}}$ possesses a $\mathbb{Z}_2 $ 1-form $\times \mathbb{Z}_2$ 1-form symmetry. The first symmetry action is generated by $S_c=\prod_{p,p\in \partial c} X_p$, a product of $X$s acting on faces belonging to the boundary of a cube, and the second symmetry action is generated by $S_v=\prod_{l,v\in \partial l} X_l$, a product of $X$s acting on links emanating from a vertex (i.e. it is the boundary of a cube in the dual lattice). $H_{\text{SPT}}$ can be obtained from a trivial paramagnet $-\sum_p X_p -\sum_l X_l$ by a depth-1 untiary $U_{\text{CZ}}=\prod_{\expval{l,p}}\text{CZ}_{l,p}$, i.e. a product of CZ gates acting on every neighboring two  qubits (one defined on a link and one defined on a plaquette). Correspondingly, $H_{\text{SPT}}$ admits a unique gapped ground state $\ket{\text{SPT}} = U_{CZ} \otimes_p \ket{+}_p\otimes_l \ket{+}_l$. More insightfully, in Pauli-X bases,  $\ket{\text{SPT}}$ can be written as two species of loop condensates with braiding structure \cite{Chong_2015_loop}:
	\begin{equation}\label{eq:3d_spt}
		\ket{\text{SPT}}	=  \sum_{ \mathcal{C}_A, \mathcal{C}_B} (-1)^{\chi( \mathcal{C}_A, \mathcal{C}_B    )} \ket{\mathcal{C}_A, \mathcal{C}_B    }, 
	\end{equation}
	where $\mathcal{C}_A$ and $\mathcal{C}_B$ are closed loops in the direct lattice (for spins on links) and the dual lattice (for spins on plaquettes), and $\chi( \mathcal{C}_A, \mathcal{C}_B    )$ counts the number of times when $\mathcal{C}_A$ braids with $\mathcal{C}_B$. Physically, two types of loop configurations can be regarded as the domain walls of two 3d Ising gauge theories (one defined on a direct lattice with the Hamiltonian $- \sum_p \prod_{ l \in \partial p} Z_l$ and the other one defined on the dual lattice with the Hamiltonian $\sum_l \prod_{p, \partial p \ni l} Z_l$). Since the loops are product states in Pauli-X bases, performing $X$ measurement on one type of spins, say all spins on links, projects them to a product state with a particular $\mathcal{C}'_A$ configuration. On the other hand, the un-measured spins exhibits a $\mathbb{Z}_2$ topological order as it is the superposition of  closed loops in the dual lattice. Specifically, the post-measurement state is 

	\begin{equation}
		\ket{\psi}  =  \ket{ \mathcal{C}'_A } \otimes \sum_{\mathcal{C}_B} (-1)^{\chi( \mathcal{C}'_A, \mathcal{C}_B    )} \ket{\mathcal{C}_B    }. 
	\end{equation}
To remove the phase $(-1)^{\chi( \mathcal{C}'_A, \mathcal{C}_B )}$, one can apply a product of Pauli-Xs acting on faces of 2d open membranes whose boundary form the domain wall configuration $\mathcal{C}'_A$. Applying the membrane operators leads to $  \ket{ \mathcal{C}'_A } \otimes \sum_{\mathcal{C}_B} \ket{\mathcal{C}_B    }$, where the un-measured species forms an exact ground state of the 3d toric code, which exhibits a $\mathbb{Z}_2$ topological order.

The above disentangling perspective can be summarized as follows. Starting from a classical model in $d$ space dimensions with $p$-form $\mathbb{Z}_2$ symmetry, there is a corresponding ``dual'' classical model in the same space dimension with $q$-form $\mathbb{Z}_2$ symmetry ($q+p=d-1$). Braiding their domain-wall configurations leads to $\ket{\text{SPT}}= \sum'_{\alpha,\beta}(-1)^{\chi(\alpha,\beta)} \ket{\alpha,\beta }$, a fixed-point SPT protected by a $p$-form $\mathbb{Z}_2\times q$-form $\mathbb{Z}_2$ symmetry, where the first and second symmetry acts on the species $A$ and $B$ respectively. It follows that measuring either species gives rise to a $\mathbb{Z}_2$ long-range (symmetry-breaking or topological) order for the other. For instance, within this framework, there are two choices for constructing two distinct SPTs in $d=3$ space dimensions. The first choice is $(p,q)=(1,1)$, and the corresponding SPT is given by two species of fluctuating loops with braiding structure, i.e. Eq.\ref{eq:3d_spt}. Alternatively, one can adopt the second choice: $(p,q)=(0,2)$, equivalent to $(p,q)=(2,0)$, and the SPT is obtained by braiding closed 2d membranes and pairs of point-like objects. In that case,  depending on which species we measure, one can obtain either a $Z_2$ spontaneous symmetry breaking order or a $\mathbb{Z}_2$ topological order of 3d toric code as a membrane condensate.

\textbf{Long-range order by measuring non-fixed-point SPT}	--  Here we show that even after deforming the  fixed-point SPT with a symmetric finite-depth unitary of the form $U_A \otimes U_B$ with $U_A, U_B $ acting on species $A,B$, measuring one species still leads to a long-range order for the un-measured species.

We start with the following observation: $\ket{\text{SPT}}$ protected by a $p$-form $\mathbb{Z}_2\times q$-form $\mathbb{Z}_2$ symmetry exhibits maximal entanglement between $A$ and $B$ in their symmetric subspaces. For instance, choosing $\ket{\text{SPT}}$ as the ground state of the model defined in Eq.\ref{eq:2d_cluster_spt}, there are $L^2$ spins defined on sites (species $A$), but the symmetric subspace of $A$ has a dimension $d=2^{L^2-1}$ due to the symmetry restriction $\prod_{v}X_v=1$. On the other hand, there are $2L^2$ spins defined on links (species $B$), but due to the symmetry restriction $\prod_{l\in \mathcal{C}} X_l=1$ ($\mathcal{C}$ is any closed loop), the dimension of the symmetric Hilbert space of $B$ is $d=2^{L^2-1}$ as well. A straightforward calculation shows that the entanglement between $A$ and $B$ is $\log d= (L^2-1)\log 2$-- maximal entanglement between  $A$ and $B$  in the symmetric subspace. This property implies that for any symmetric unitary $U_A$ acting on $A$, there always exists a symmetric unitary  $V_B$ acting on $B$ such that  $U_A\ket{\text{SPT}} =V_B \ket{\text{SPT}}$ (see Appendix.\ref{appendix:spt_push} for details). Moreover, since $U_A$ is a finite-depth local unitary, so is $V_B$. Therefore, measuring species $A$ via the projector $P_A$ onto the state $|\alpha'\rangle$ gives $P_A (U_A \otimes U_B) \ket{\text{SPT}} = P_A U_B V_B \ket{\text{SPT}}=  \ket{\alpha'} \otimes U_B V_B\sum'_{\beta} (-1)^{\chi(\alpha',\beta) } \ket{\beta} $ because $P_A U_B V_B=U_B V_B P_A$.  Since $U_B V_B$ is a symmetric finite depth circuit, $U_B V_B\sum'_{\beta} (-1)^{\chi(\alpha',\beta) } \ket{\beta} $ is in the same phase as $\sum'_{\beta} (-1)^{\chi(\alpha',\beta) } \ket{\beta}$ and thus exhibits the same long-range order.

\subsection{From SPT to SET}
The disentangling perspective provides a simple way to prepare symmetry-enriched topological (SET) order with constant-depth local adaptive circuits. As discussed in Ref.\cite{mcgreevy_2016_snake}, one can decorate the fluctuating loops in Eq.\ref{eq:3d_spt} with 1d SPTs by introducing extra spins defined on sites of the direct lattice and the dual lattice. This construction leads to the following state 

\begin{equation}\label{eq:decorated_spt}
	\ket{\psi_0}  =    \sum_{\mathcal{C}_A, \mathcal{C}_B} (-1)^{\chi( \mathcal{C}_A, \mathcal{C}_B    )} \ket{\mathcal{C}_A,\mathcal{C}_B    } U_{\mathcal{C}_A} \ket{+}^{\otimes v} U_{\mathcal{C}_B} \ket{+}^{\otimes \tilde{v}}
\end{equation}
where $U_{\mathcal{C}_A}$ ($U_{\mathcal{C}_B}$) creates a 1d SPT along the loop $\mathcal{C}_A$ ($\mathcal{C}_B$) for the extra spins defined on sites of the direct (dual) lattice labeled by $v$ ($\tilde{v}$).  Choosing $U_{\mathcal{C}_A } = \prod_{\expval{v,v'} \in \mathcal{C}_A } CZ_{\expval{v,v'}}$ creates a 1d SPT with $\mathbb{Z}_2 \times  \mathbb{Z}_2$ symmetry. Specifically, $\ket{\psi_0}$ can be prepared as follows: starting from Eq.\ref{eq:3d_spt}, we  initiate extra spins in $\ket{+}$, i.e. the $+1$ eigenstate of Pauli-X. For each link $l$ and its two boundary sites $(v,v')$ in the direct lattice, we consider the unitary gate $u_l=(\ket{+}\bra{+})_l \otimes \mathbb{I}_{\expval{v,v'}} + (\ket{-}\bra{-})_l \otimes CZ_{\expval{v,v'}}$, i.e. a gate that can be obtained by conjugating a $CCZ$ gate with a Hadarmard gate on link $l$. Similarly, one can define $u_{\tilde{l}}$ acting on a link $\tilde{l}$ and its two boundary sites in the dual lattice. Simultaneously applying $u_l$ ($u_{\tilde{l}}$) on every link in the direct (dual) lattice leads to $\ket{\psi_0} $ defined in Eq.\ref{eq:decorated_spt}. Due to the decoration, in addition to two $\mathbb{Z}_2$ 1-form symmetries, $\ket{\psi_0}$ also respects two global $\mathbb{Z}_2$ symmetries acting on sites of the  direct (dual) lattice. 

Measuring Pauli-X for spins on the links leads to the state 
\begin{equation} 
\left(\ket{\mathcal{C}'_A}  \otimes U_{\mathcal{C}'_A} \ket{+}^{\otimes v} \right) \otimes \left(\sum_{ \mathcal{C}_B} (-1)^{\chi( \mathcal{C}'_A, \mathcal{C}_B    )} \ket{\mathcal{C}_B    }  U_{\mathcal{C}_B} \ket{+}^{\otimes \tilde{v}}\right).
\end{equation}

By applying a product of Pauli-Xs to remove the phase factor $(-1)^{\chi(\mathcal{C}'_A , \mathcal{C}_B    )}$ as discussed in the previous section, one obtains  a superposition of loops $\mathcal{C}_B$ in the dual lattice, where each loop is further decorated by a 1d SPT with $\mathbb{Z}_2\times \mathbb{Z}_2 $ symmetry: 

\begin{equation}
\ket{\text{SET}} =  \sum_{ \mathcal{C}_B}  \ket{\mathcal{C}_B    }    \otimes  U_{\mathcal{C}_B} \ket{+}^{\otimes \tilde{v}}. 	
\end{equation}

Notably, such a state exhibits an SET order, as (deconfined)  anyonic excitations appear in conjunction with boundaries of  1d SPTs, thereby exhibiting a further symmetry fractionalization (see Ref.\cite{mcgreevy_2016_snake} for more details).

\subsection{Preparing CSS codes within chain complex description}
In the above discussion, the SRE states arise by braiding two species of fluctuating domain walls of a local classical model and its dual. In Appendix.\ref{appendix:disentangling_css}, we relax the geometric locality by considering the domain walls of $k$-local classical models within the framework of chain complex. This allows us to construct SRE states without spatial locality. By measuring one species of domain walls, one can obtain any CSS code characterized by a chain complex.

\section{Summary and discussion}

In this work, we exploit measurement to construct four distinct classes of local adaptive circuits. This allows for a low-depth preparation for various long-range entangled states characterized by gapped topological orders and gapless CFTs. Below we summarize each class of adaptive protocols and present open questions regarding the power of local adaptive circuits.

Our first protocol exploits the tensor network representation of topologically ordered states. By preparing copies of constant-size states, each of which respects a given local symmetry, performing measurement glues these copies into a many-body state with local gauge symmetry, thereby exhibiting certain topological orders. For preparing quantum double models of finite abelian groups, correcting measurement errors can be regarded as annihilating topological defects (i.e. abelian anyons), which can be implemented in constant time. However, anyons in non-abelian quantum double cannot be  annihilated via a finite-depth circuit, and it remains to be understood whether the tensor-network-based adaptive protocol can prepare this class of states or not.  Nonetheless, for preparing abelian topological orders, this modular assembly is practical for photonic and trapped ion systems.

Our second protocol provides a framework for realizing MERA circuits using local unitaries and measurements.  This allows the efficient preparation of 1d gapless critical states characterized by 1+1D conformal field theories in $O(\log L )$ depth. Notably, our preparation scheme is optimal since, as we proved, starting from a product state, any depth-$D$ local adaptive circuit can \Lu{at most} generate entanglement of  $O(D|\partial A|)$ with $|\partial A|$ being the area of a subregion $A$. However, as CFTs in higher space dimensions satisfy an area-law bound of entanglement, it is natural to ask whether there exists a finite-depth adaptive circuit for preparing them. If the answer is negative, is there any fundamental obstruction?

Our third protocol is inspired by parton constructions, where physical degrees of freedom are split into partons subject to certain constraints. This motivates the protocol of first preparing a short-range entangled state of partons, and then performing measurements to enforce the parton constraints, thereby obtaining long-range entangled states. While we only provide a detailed construction for the Kitaev honeycomb model, we expect it can be readily generalized to other parton-inspired adaptive circuits. Moreover, even without correcting measurement outcomes, it would be interesting to explore the possibility of unconventional (disordered) quantum states of matter due to measurement.

Finally, building on \cite{ashvin_2021_measurement, verresen2021_measurement_cold_atom}, we consider certain short-range entangled states that describe two species of fluctuating $\mathbb{Z}_2$ domain walls with non-trivial braiding structures. Performing measurement disentangles the two species; the measured species becomes a trivial product state and the other species forms a state of fluctuating domain walls, hence possessing long-range order. This idea leads to a simple understanding of emergent topological orders and their symmetry-enriched versions by measuring certain SPT fixed-point wave functions. Moreover, we prove that the emergence of topological orders upon measurements is a property shared by certain separable finite-depth symmetric unitary deformations of the fixed-point SPTs. Whether the emergence of topological order via measurement is a universal property of the SPT phase remains an open question.

In general, it is an open question to determine all constraints and bounds for local adaptive circuits, which would shed light on which states can or cannot be prepared efficiently.  The Lieb-Robinson bound is clearly violated due to the measurements, but local adaptive circuits have bounded entanglement growth (Appendix \ref{appendix:entanglement_bound}); are there other restrictions?  Recent progress has been made \cite{adaptivebounds} toward addressing this question.  It is also interesting to compare the power of local adaptive circuits with the power of spatially non-local (but $k$-local) unitary circuits.  The two classes certainly overlap (our MERA protocol illustrates this), but generally differ in the depth required to prepare states.  For instance, $k$-local circuits can generate volume law entanglement in finite time (while local adaptive circuits cannot), whereas local adaptive circuits can prepare GHZ and other long-range entangled states in finite time (while $k$-local circuits require at least $O(\log L)$ depth).

	%------------------------------------------------------------

	\acknowledgements{We thank Tarun Grover, Yin-Chen He, John McGreevy, Nathanan Tantivasadakarn, Beni Yoshida, Yi-Zhuang You, and Liujun Zou for useful discussions, and Gabriela Secara for assistance with figures. We also thank Ruben Verresen, Nathanan Tantivasadakarn, Ryan Thorngren, Ashvin Vishwanath for email correspondence. IK thanks Sergey Bravyi, Alexander Kliesch, and Robert Koenig for helpful discussions and collaboration on a related work. This work was supported by Perimeter Institute and NSERC.
Research at Perimeter Institute is supported in part by the Government of Canada through the Department of Innovation, Science and Economic Development Canada and by the Province of Ontario through the Ministry of Colleges and Universities.
	}
	
	\bibliography{v4bib}
	
	%------------------------------------------------------------
	\newpage 
	\appendix

%\onecolumngrid

\section{Adaptive circuits via tensor-networks}

\subsection{ $\mathbb{Z}_2$ topological order away from fixed point}\label{appendix:non_fixed_point}
Here we present the details for preparing non-fixed-point states with $\mathbb{Z}_2$ topological order, motivated by the tensor-network construction in Ref.\cite{wen_tn_symmetry_2010}. Consider the same tensor-network structure as for the toric code (Fig.\ref{fig:1}a), one defines a perturbed $T$ tensor: $T(p)_{\alpha\beta\gamma\delta }  =  p^{ - (\alpha+ \beta+ \gamma +\delta)  }\delta_{ \alpha+ \beta+ \gamma +\delta, 0}  $. Note that the delta function is responsible for the closed-loop condition, where $ \alpha+ \beta+ \gamma +\delta$ is defined mod 2.  While $p=1$ gives the exact toric code state, any $p>1$ imposes a string tension so that the weight of a string segment is reduced by a factor of $p^{-2}$. The corresponding tensor network state is a superposition of loops with amplitudes decaying exponentially with the length of the loops: 

\begin{equation}\label{eq:perturbed_toric}
\ket{ \psi_0(p)}  =  \sum_{\mathcal{C}}  p^{ - 2\abs{\mathcal{C}} } \ket{\mathcal{C}},
\end{equation}
where $\abs{\mathcal{C}}$ is the string length of the closed loop $\mathcal{C}$. The state $\ket{\psi_0(p)}$ is a ground state of a frustration-free Hamiltonian (see below for its construction), and it exhibits the $\mathbb{Z}_2$ topological order up to a finite critical $p_c$. This can be seen from the norm  of the un-normalized state $\braket{\psi_0 (p)}{\psi_0(p)} =\sum_{\mathcal{C}}  p^{-4\abs{\mathcal{C}}} $, which is exactly the thermal partition function of the 2d Ising model using a high-temperature expansion, i.e.  $Z\sim \sum_{\mathcal{C}} \tanh( \beta)^{\abs{\mathcal{C}}}$. Alternatively, the stability of topological order has been confirmed by numerically computing the topological entanglement entropy \cite{wen_tn_symmetry_2010}.

The above tensor-network construction motivates the following  local adaptive circuit. First, using the $T(p)$ tensor, we construct a local vertex state of four spins $\ket{T(p)}_v = \sum_{\alpha,\beta,\gamma,\delta}T(p)_{\alpha \beta\gamma \delta}    \ket{ \alpha,\beta,\gamma,\delta }$. Notice that it can be written as 

\begin{equation}\label{eq:perturbed_vertex}
\ket{T(p) }_v  = p^{ -\sum^{v}  \frac{1-Z}{2}  }  \ket{ T(p=1)}_v,  
\end{equation}
where $\sum^v$ denotes a summation over the four spins, and $\ket{T(p=1)}_v$ is a stabilizer state with  $ZZZZ$ stabilizer and three independent $XX$ stabilizers, say $X_1X_2, X_2X_3, X_3X_4$. For $p>1$, while $ZZZZ$ remains a symmetry, i.e. $ZZZZ \ket{T(p)} = \ket{T(p)}$, $XX$ is no longer a symmetry. Preparing the vertex state $\ket{T(p)}_v$ on every vertex of a 2d square lattice, we perform a two-body $ZZ$ measurement for  two spins on the same link, and with the measurement outcome $ZZ=1$ on every link, one obtains the state
\begin{equation}
\ket{\psi(p)}	 = \left[ \prod_l  \frac{1+  \prod_{ v \in \partial l }Z_{l,v}   }{2} \right] \otimes_v \ket{ T(p)}_v, 
\end{equation}
In the subspace fixed by $ZZ=1$, one has an effective qubit on every link, and by construction, the effective qubits exactly form the state given by Eq.\ref{eq:perturbed_toric}, which therefore exhibits a topological order up to $p_c$.

\begin{figure}
	\centering
	\begin{subfigure}{0.40\textwidth}
		\includegraphics[width=\textwidth]{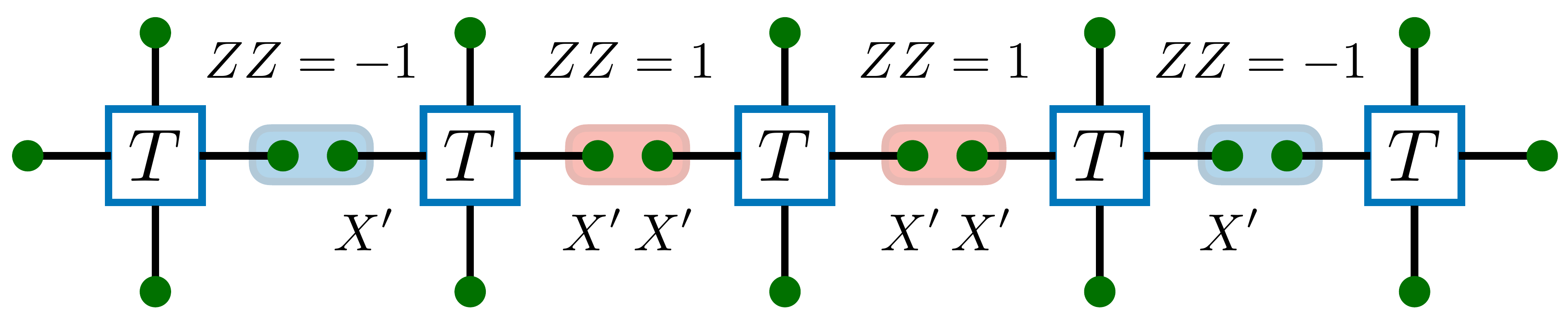}
	\end{subfigure}
	\caption{The measurement outcome $ZZ=-1$ on two links can be corrected by applying a non-unitary string operator consisting of onsite operator $X'\equiv p^{ Z} X $.}
	\label{fig:2dtoric_perturbed_correction}  
\end{figure}

Same as the exact toric code, the measurement error, i.e. $ZZ=-1$, comes in pairs since the product of all $ZZ$ measurement operators on links is the product of all $ZZZZ$ stabilizers of vertex states, thereby a symmetry of the state $\otimes_v \ket{ T(p)}_v$. To correct any two errors, one requires  an operator that anticommutes with those two $ZZ$ operators, but meanwhile, acts trivially on $\otimes_v \ket{T(p)}_v$. It turns out such an operator is a product of $X'=p^{ Z} X  = e^{ Z \log p } X $ along a string that connects two  errors (see Fig.\ref{fig:2dtoric_perturbed_correction}).  One caveat is  that such a string operator is non-unitary (and also non-hermitian). However, we note that in the subspace fixed by a given set of measurement outcomes, the post-measurement state is equivalent to

\begin{equation*}
	\ket{\psi (p)} = e^{\sum_l m_l \tilde{Z}_l } \ket{\mathcal{T}   },
\end{equation*}
where $\ket{\mathcal{T} }=\sum_{\mathcal{C}} \ket{\mathcal{C}}$ is the toric code state, $\tilde{Z}_l$ is the Pauli-Z operator acting on the effective qubit per link, and $m_l \in \{ -1, 0,1 \}$ is  fixed by both (i) the measurement outcomes and (ii) how we define the effective qubits. Specifically, $m_l=0$ at the location of $-1$ measurement outcomes (i.e. errors). Since these errors occur in  pairs, they correspond to the boundary of open strings $\Gamma$, and one assigns $m_l = -1$ along $\Gamma$ except for their boundary. Finally, $m_l$ is $1$ for those links that are neither in the interior of those strings nor at the location of errors. Physically, $m_l = -1,  0, 1$ corresponds respectively to assigning negative, zero, and positive string tension on the toric code. Since topological order is generically stable in the presence of weak string tension, one expects $\ket{\psi(p)}$ to be topologically ordered up to a certain $p_c\geq 1$.

\noindent\textbf{Derivation of parent Hamiltonian}: Here we derive a parent Hamiltonian for the state defined in Eq.\ref{eq:perturbed_toric} using an approach analogous to Ref.\cite{rg_mcgreevy_2016,topo_scar_2019_chamon}. Define the four-body operator around every plaquette $p$: $Q_p =  p^{-2 \sum_{l\in \partial p} Z_l    } -  \prod_{ l \in \partial p } X_l $, by writing $\ket{\psi_0(p)}\sim  p^{ \sum_l Z_l} \sum_{\mathcal{C}} \ket{\mathcal{C}}$, it is straightforward to see that $\prod_{ l \in \partial p } X_l   \ket{ \psi_0(p)}  =p^{-2 \sum_{l\in \partial p} Z_l    } \ket{\psi_0 (p)}  $, namely, $Q_p \ket{ \psi_0(p)}=0$. On the other hand, $Q_p$ is a positive semi-definite operator by observing that
\begin{equation}
Q_p^2= Q_p \left( p^{ 2 \sum_{l\in \partial p} Z_l    }  +  p^{ -2 \sum_{l\in \partial p} Z_l    } \right). 	
\end{equation}
Therefore, $\ket{\psi_0(p)}$ lies in the zero-energy subspace of the following Hamiltonian 
\begin{equation}
\sum_{p}	 Q_p =  \sum_p \left(  p^{-2 \sum_{l\in \partial p} Z_l    } -  \prod_{ l \in \partial p } X_l   \right). 
\end{equation}
Since the state $\ket{\psi_0(p)}$  satisfies the constraint that four Pauli-Z on links emanating from a vertex $v$ is fixed at one, one can impose it energetically to derive the exact Parent Hamiltonian whose unique ground state of zero energy is $\ket{\psi_0(p)}$: 

\begin{equation}
H=   \sum_{v } \left(1-\prod_{l,  \partial l \ni v }  Z_l  \right) + \sum_p \left(  p^{-2 \sum_{l\in \partial p} Z_l    } -  \prod_{ l \in \partial p } X_l   \right). 
\end{equation} 
Notice that while the vertex terms commute with the plaquette term, neighboring plaquette terms do not commute. $H$ is therefore frustration-free but not a commuting-projector Hamiltonian.

\subsection{Quantum double models}\label{appendix:qd}
Here we discuss a local adaptive circuit for  preparing quantum double of finite Abelian group $G$. To define a quantum double \cite{kitaev2003fault}, we consider an oriented 2d lattice, where spins are defined on links, and the Hilbert space of each spin is spanned by the orthogonal basis states $\ket{g}$ with $g$ being the elements in a finite group $G$. Note that reversing the orientation of a link specified by a group element $g$ is equivalent to considering the inverse of $g$:

\begin{equation}
\includegraphics[width=5cm]{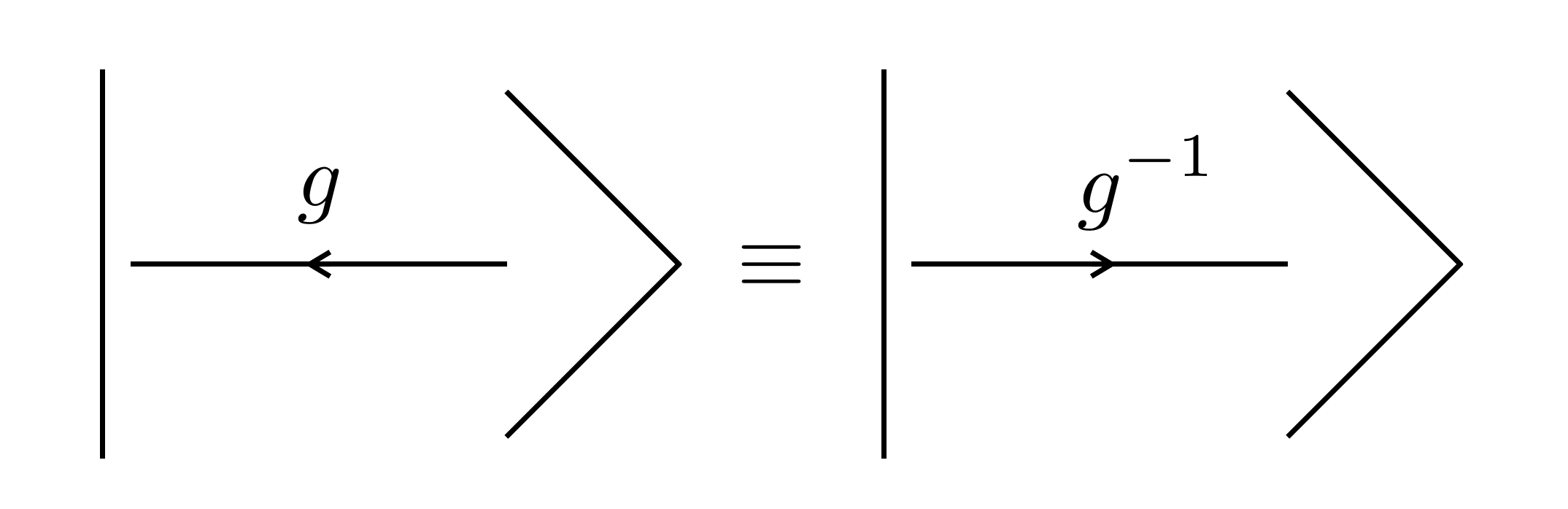}
\end{equation}
The total Hilbert space is a tensor product of local Hilbert spaces on links. One defines the vertex operators $A_v (h)$ that  perform the local gauge transformation and the plaquette operators $B_p$ that enforce  the flux-free condition:

\begin{equation}
\includegraphics[width=8cm]{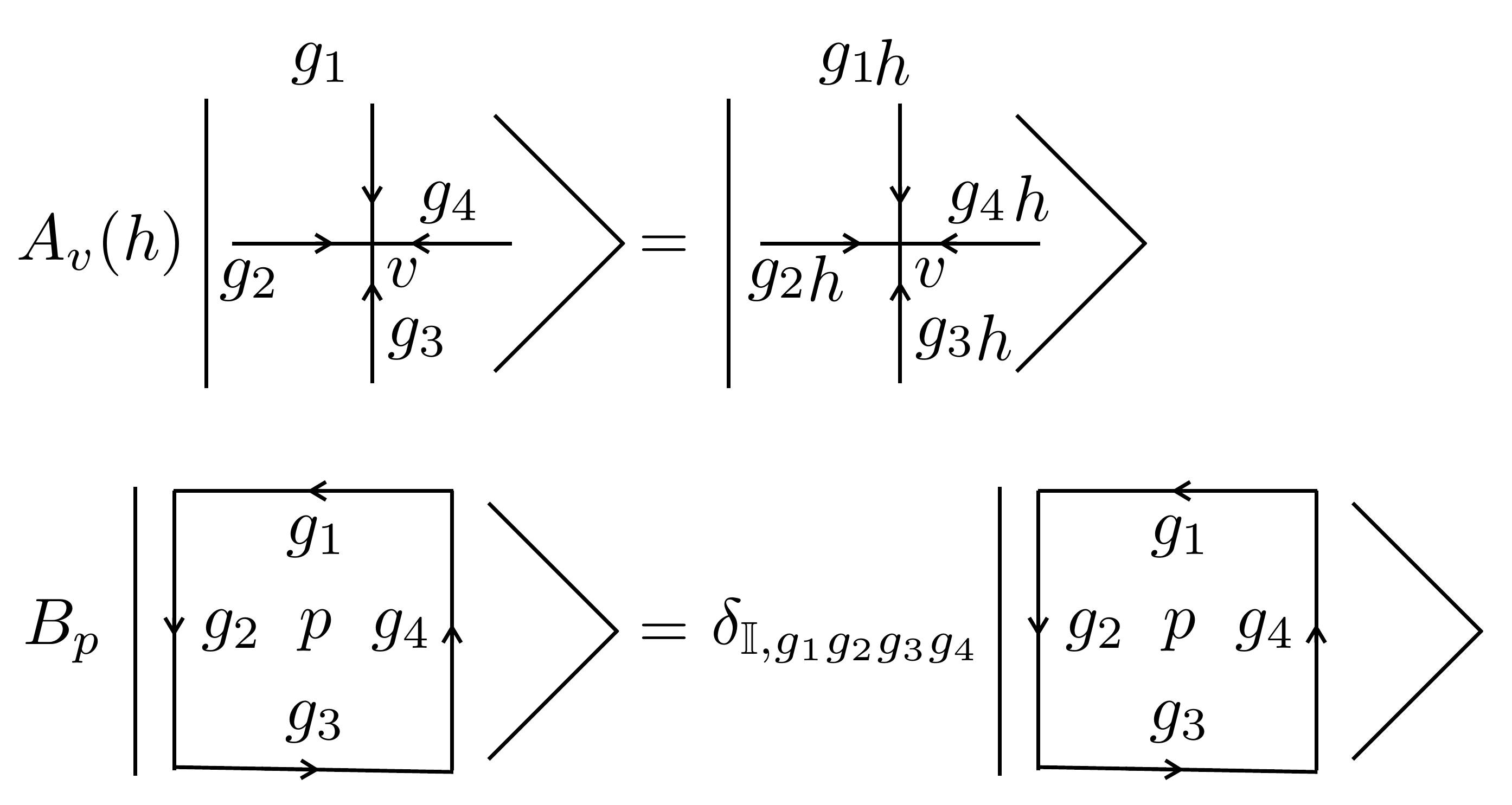}
\end{equation}
Introducing the vertex operator $A_v = \frac{1}{\abs{G}}\sum_{h\in G } A_v(h)$, which is a projector to the gauge-invariant subspace, the Hamiltonian of the quantum double is
\begin{equation}
H =  -  \sum_v A_v - \sum_p B_p, 
\end{equation}
which is exactly solvable as all operators commute: $[A_v,A_{v'}]= [B_p, B_{p'}] = [A_v , B_p]=0$, and the ground subspace is specified by $A_v =B_p =1$ for all vertices $v$ and plaquette $p$. 

Here we discuss the finite-depth adaptive circuit for preparing a quantum double ground  state $\ket{\psi}$. To start, since the vertex operator $A_v$ is not diagonal in the computational bases $\ket{g}$, it is more convenient to consider the dual lattice so that the plaquette operator $B_p$  can be regarded  as a vertex operator $\tilde{A}_{\tilde{v}}$, which is  diagonal in the computational bases:

\begin{equation}
\includegraphics[width=8.7cm]{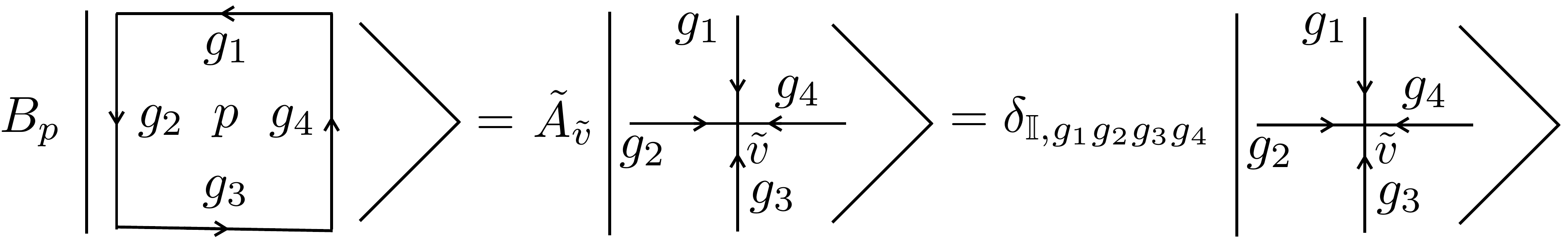}
\end{equation}

In the dual lattice, one  defines a local vertex state $\ket{T}_{\tilde{v}}= \tilde{A}_{\tilde{v}}\sum_{g_1,g_2,g_3,g_4} \ket{ g_1,g_2,g_3,g_4  } = \sum_{g_1,g_2,g_3,g_4} \delta_{ \mathbb{I}, g_1 g_2 g_3 g_4  }  \ket{ g_1,g_2,g_3,g_4  }$.  Taking a tensor product over these vertex  states: $\otimes_{\tilde{v}} \ket{T}_{\tilde{v}}$, we perform the two-body forced measurement to fuse the two spins on the same link via the projector $P_l= \sum_g  \ket{gg} \bra{gg}  $, giving the post-measurement state

\begin{equation}
\ket{ \psi }  = \left( \prod_l   P_l   \right) \otimes_{\tilde{v}} \ket{T}_{\tilde{v}},
\end{equation}
which exhibits  the topological order of quantum double.

\begin{figure}
	\centering
	\begin{subfigure}{0.47\textwidth}
		\includegraphics[width=\textwidth]{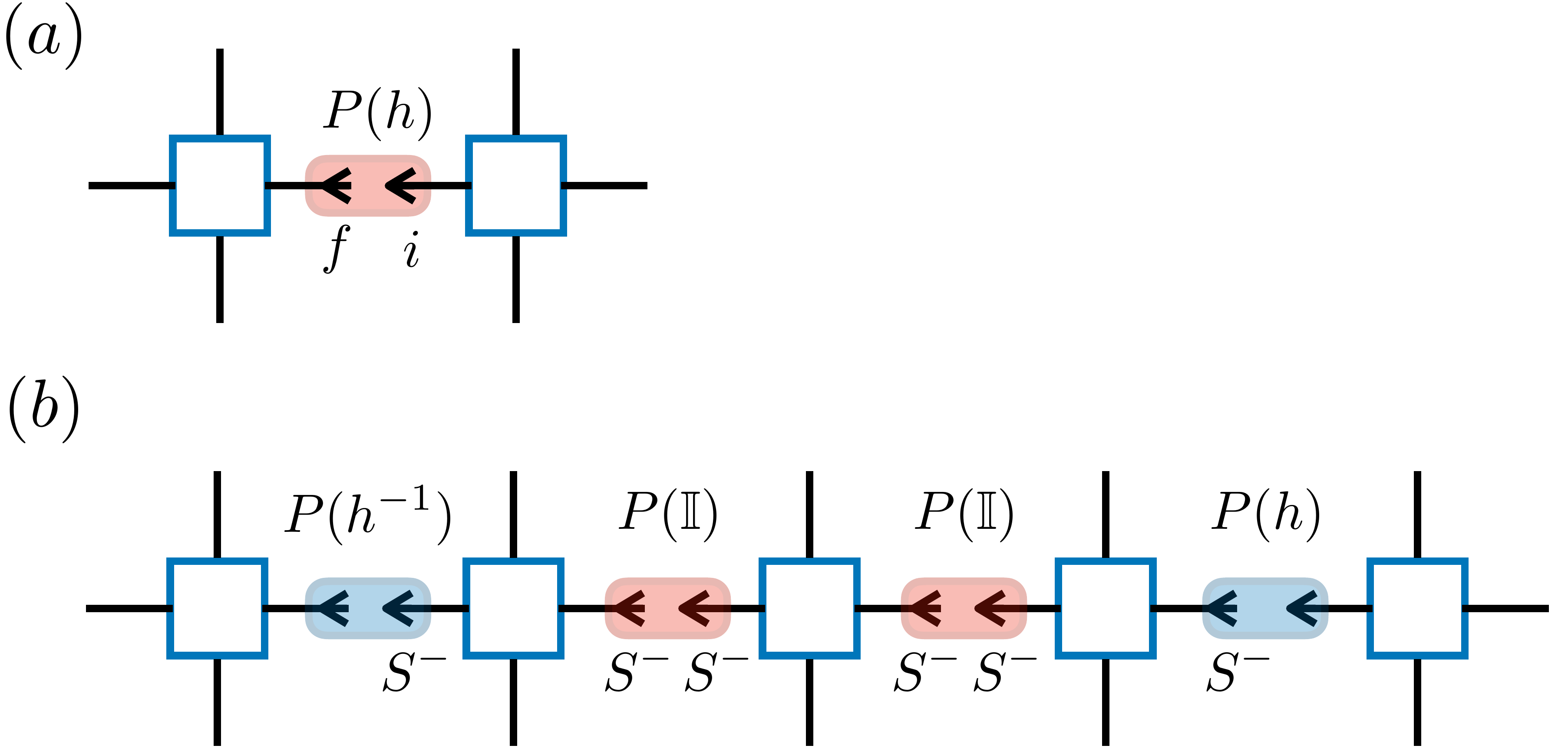}
	\end{subfigure}
	\caption{(a) The two-body projector $P(h) $ defined as $P(h)=\sum_g  \left( \ket{g} \bra{g} \right)_i  \otimes    \left( \ket{gh} \bra{gh} \right)_f =      \sum_{ g\in G}  \ket{ g,gh}\bra{g,gh}$. (b) Unwanted measurement outcomes (colored in blue) can be corrected by a string operator consisting of the product of $S^- = S^-(h)$.}
	\label{fig:qd_correction}  
\end{figure}

\begin{figure*}
	\centering
	\begin{subfigure}{0.95\textwidth}
		\includegraphics[width=\textwidth]{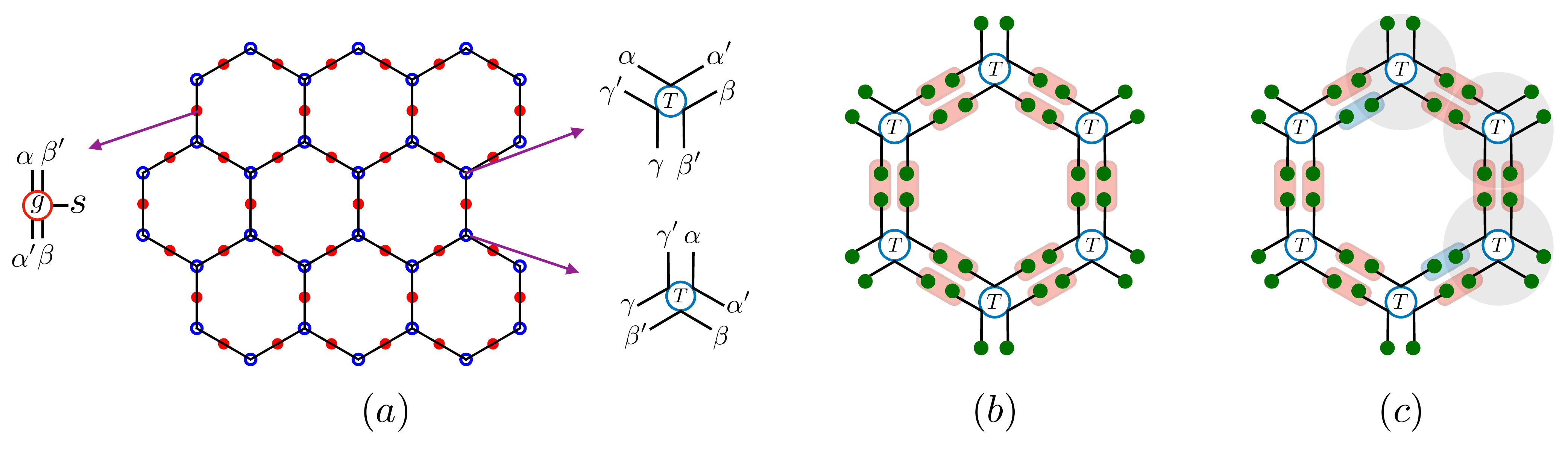}
	\end{subfigure}
	\caption{Local adaptive circuits for preparing double semion $\ket{\text{DS}}$: (a) Tensor-network representation of $\ket{\text{DS}}$ built from the tensors $T$ and $g$ (defined in Eqs.\ref{eq:ds_T},\ref{eq:ds_g}). (b) By preparing decoupled vertex states using $T$ tensor,  simultaneously measuring two-body $ZZ$ operators (colored in red) results in the double semion state $\ket{\text{DS}}$ when the measurement outcome is $ZZ=1$. (c) Measurement errors ($ZZ=-1$, colored in blue) come in pairs on the inner loop of each plaquette. Errors can be corrected by simultaneously applying a product of X-type operators (defined in Eq.\ref{eq:ds_tn_operator}) that connect them, where each X-type operator (colored in gray) contains two Pauli-Xs acting on the inner loop.}
	\label{fig:ds}  
\end{figure*}

Now we discuss how to correct unwanted measurement outcomes, which is only applicable to quantum double with a finite Abelian group. First, we define a measurement operator $M$ that encodes all measurement outcomes and the corresponding projected subspaces. To this end, we  define a projector $P(h)= \sum_{ g\in G}  \ket{ g,gh}\bra{g,gh}$ for $h\in G$. One can check that $\sum_{h \in G} P(h)=1$. Now one can introduce a set of distinct real numbers $\{  \lambda_h| h\in G \}$ for the measurement outcome, and it follows that the measurement (Hermitian) operator can be written as 
\begin{equation}
M 	 = \sum_{h\in G}   \lambda_h P(h )  =\sum_h \lambda_h  \left[ \sum_{ g\in G}  \ket{ g,gh}\bra{g,gh}\right]. 
\end{equation}
The desired fusion  corresponds to the measurement outcome $\lambda_{\mathbb{I}   }$ with the projector $P= P(\mathbb{I}) =  \sum_g  \ket{gg} \bra{gg} $. Due to the local gauge constraint on each vertex $\tilde{A}_{\tilde{v}}=1$, the unwanted measurement outcome must come in pair with the projectors $P(h)$ and $P(h^{-1})$. Similar to the 2d toric code, the errors on two links can be corrected by applying a string operator consisting  of $S^-(h)$ with  $S^-(h)\ket{g} = \ket{ gh^{-1} } $ (see Fig.\ref{fig:qd_correction}).

\subsection{Double semion}\label{appendix:ds}
We here discuss the preparation for the double semion ground state \cite{wen_string_net2005}. Consider a honeycomb lattice with each link accommodating a qubit, the state is a superposition of product states of loops weighted by a sign that depends on the number of loops in $\mathcal{C}$: $ \ket{\text{DS}}	 = \sum_\mathcal{C} (-1)^{N_\mathcal{C}} \ket{\mathcal{C}}$. $\ket{\text{DS}}$ admits an exact tensor-network representations \cite{wen_tnrg_2008,tn_string_net_wen_2009}  (see Fig.\ref{fig:ds}a):

\begin{equation}
\ket{ \text{DS}  }   =  \sum_{\{s_i\}}    \text{tTr}  \left[   \otimes_v T \otimes_l g^{s_i}   \right]  \ket{ \{s_i\} }
\end{equation}
where one defines  the 6-leg tensor $T$ on vertices and the 5-leg  tensor $g$ on links as follows:  

\begin{equation}\label{eq:ds_T}
\begin{split} &T_{\alpha\alpha';\beta\beta';\gamma\gamma'}  =  ~T^0_{\alpha\beta \gamma} \delta_{\alpha \alpha'} \delta_{\beta \beta'} \delta_{\gamma\gamma'}  \\\\[1pt]
& T_{\alpha \beta \gamma}^0 =  \begin{cases}
1 ~\text{   if  } \alpha+\beta +\gamma =0,3 \\
i ~\text{   if  } \alpha+\beta +\gamma =1		 		\\ 
-i ~\text{   if  } \alpha+\beta +\gamma =2, 	
\end{cases}
\end{split}
\end{equation}
and 
\begin{equation}\label{eq:ds_g}
g^{s}_{\alpha\beta ;  \alpha' \beta '}	  =  \delta_{ s, \alpha  +  \beta   }  \delta_{\alpha \alpha' } \delta_{\beta \beta' }.  
\end{equation}
Essentially, every link has a structure of double lines, each of which is associated with a virtual qubit. The physical qubit on a link takes the value 0/1 when the two virtual qubits take the same/opposite value. In addition, since all the virtual qubits on a line surrounding a hexagon take the same value, one may view those virtual qubits as an Ising spin located at the center of a hexagon, and the physical qubit on a link simply measures the domain wall between two Ising spins belonging to two neighboring hexagons. On the other hand, the non-trivial phase from $T_{\alpha\beta \gamma}^0$  provides the correct sign associated with each loop configuration in the double semion state.

The above tensor-network construction motivates the following adaptive circuit. For each vertex, we employ the $T$ tensor to define a vertex state of 6 physical spins: 
\begin{equation}
\ket{T}_v  = \sum_{  \alpha,\alpha',\beta,\beta',\gamma,\gamma'} T_{\alpha\alpha';\beta\beta';\gamma\gamma'}  \ket{  \alpha,\alpha',\beta,\beta',\gamma,\gamma' },
\end{equation}
where we note that $\ket{T}_v$ is the common eigenstate with eigenvalue $1$ for the following six (mutually commuting) operators 
\begin{equation}\label{eq:ds_tn_operator}
\begin{split}
\{ &Z_\alpha Z_{\alpha'}, Z_{\beta } Z_{\beta'}, Z_{\gamma} Z_{\gamma'}, \\
&\sqrt{Z_{\alpha}Z_{\beta }  Z_{\gamma}  }  X_{ \alpha } X_{\alpha'}  \sqrt{Z_\alpha}   Z_{\alpha}Z_{\beta }  Z_{\gamma}, \\
& \sqrt{Z_{\alpha}Z_{\beta }  Z_{\gamma}  }  X_{ \beta } X_{\beta'}  \sqrt{Z_\beta}   Z_{\alpha}Z_{\beta }  Z_{\gamma},\\
&   \sqrt{Z_{\alpha}Z_{\beta }  Z_{\gamma}  }  X_{ \gamma } X_{\gamma'}  \sqrt{Z_\gamma}   Z_{\alpha}Z_{\beta }  Z_{\gamma}\}	
\end{split}	
\end{equation}
As we will soon discuss, identifying these operators is crucial for correcting measurement errors. Taking a tensor product of the vertex states gives $\ket{\psi_0 } = \otimes_v \ket{ T}_v$, where there are four spins on each link. We measure $ZZ$ operators of two spins on the same side of the double line (see Fig.\ref{fig:ds}b), and with the outcomes $ZZ=1$ for all measurements, one obtains  the state
\begin{equation}
\ket{\psi }	 = \left[ \prod_{ l  } \frac{1 +ZZ   }{2} \frac{1+ZZ}{2}   \right]\otimes_v \ket{ T}_v.
\end{equation}
In the subspace fixed by $ZZ=1$, the effective local Hilbert space on each link is 4 dimensional, corresponding to two effective qubits (one qubit on each side of a link). As in the tensor-network construction where a physical spin reflects a domain wall between two virtual spins located on two sides of a link, $\ket{\psi}$ exhibits the $\mathbb{Z}_2$ double semion order which is encoded in the domain-wall variables from two effective qubits on links.

Any  measurement errors ($ZZ=-1$) can be corrected via a depth-1 local unitary circuit. To start, since the product of all $ZZ$ measurement operators around each inner plaquette loop is fixed at one, the measurement errors come in pairs. This suggests that these errors can be corrected by applying a string operator $S$ as in the case for 2d toric code. Such a string operator $S$ must anticommute with the $ZZ$ operators where $-1$ outcome occurs, while acts trivially on the tensor product state $\otimes_v  \ket{T}_v$, i.e. $S\otimes_v  \ket{T}_v = \otimes_v  \ket{T}_v$. As a result, $S$ is a product of X-type operators (i.e. those operators involving Pauli-Xs in Eq.\ref{eq:ds_tn_operator}) along a path that connects two errors (see Fig.\ref{fig:ds}.c).

\subsection{X-cube fracton order}\label{sec:fracton}

\begin{figure}
	\centering
	\begin{subfigure}{0.32\textwidth}
		\includegraphics[width=\textwidth]{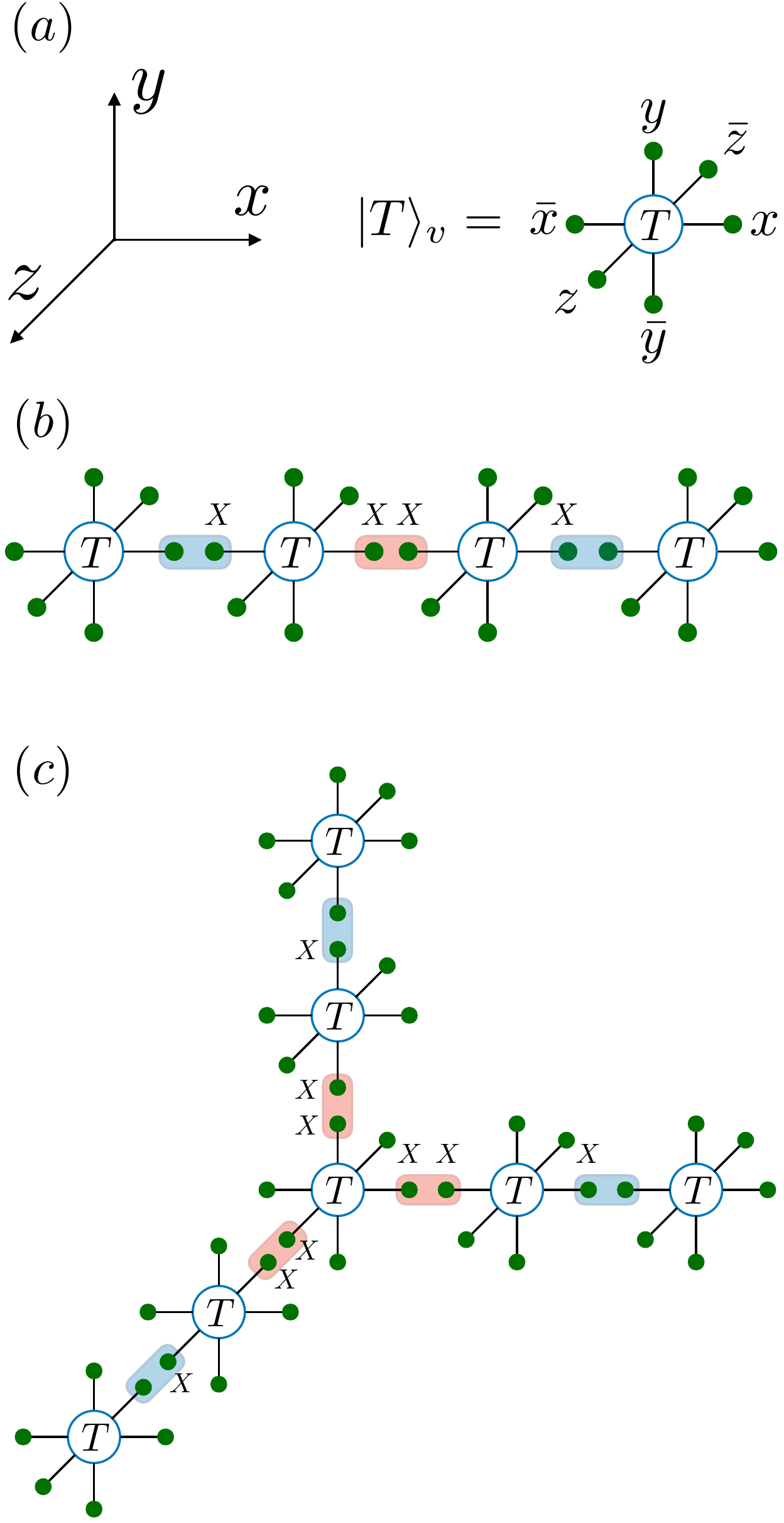}
	\end{subfigure}
	\caption{(a) We first prepare a vertex state $\ket{T}_v$ of 6 qubits (Eqs.\ref{eq:x_cube_T}, \ref{eq:x_cube_vertex}) on each vertex of a 3d cubic lattice, and measure the two-body $ZZ$ operators for two qubits on the same link. Measurement outcomes $ZZ=1$ lead to a ground state of the X-cube model (Eq.\ref{eq:x_cube}). (b) and (c) show how to correct the measurement errors ($ZZ=-1$, represented by blue rectangles). Desired outcomes $ZZ=1$ are represented by red rectangles. In (b), one applies a product of $X$ along a 1d line to annihilate two errors. In (c), one applies a product of $X$ along three lines that meet in a vertex to annihilate three errors along distinct spatial directions.}
	\label{fig_fracton}  
\end{figure}

Here we discuss the constant-depth preparation of X-cube model \cite{xcube}, a canonical example of type-I fracton topological order. To define the model, we consider a 3d cubic lattice (with periodic boundary conditions in all spatial directions), where every link accommodates a  qubit. The Hamiltonian is defined as

\begin{equation}\label{eq:x_cube}
	H= - \sum_c B_c  - \sum_{ v} \left(  A_{v,xy}+A_{v,yz}+A_{v,zx} \right),
\end{equation}
The cube term $B_c$ is a product of Pauli-Xs on 12 links around a cube $c$.  There are three types of vertex terms $ A_{v,xy}, A_{v,yz}, A_{v,zx}$, each of which is a product of four Pauli-Zs around a vertex $v$  on a given x-y, y-z, or z-x plane.	

As discussed in Ref.\cite{he_2018_stabilizer},  one can exactly represent a ground state of the X-cube model via tensor-network. To start,  on each link, one defines a 3-leg tensor $g^s_{ij}= \delta_{s,i}\delta_{s,j}$, which projects the physical index $s$ to two neighboring virtual legs $i, j$. In addition, one defines a 6-(virtual) leg tensor $T$ on each vertex:
\begin{equation}\label{eq:x_cube_T}
	T_{x\bar{x}, y\bar{y},z\bar{z}}= \begin{cases}
		1 \quad \text{if}  \begin{cases}
			x+ \bar{x}+  y+ \bar{y} = 0 \text{ mod } 2\\
			y+ \bar{y}+  z+ \bar{z} = 0 \text{ mod } 2\\
			z+ \bar{z}+  x+ \bar{x} = 0 \text{ mod } 2
		\end{cases}\\
		0\quad  \text{otherwise}.
	\end{cases}
\end{equation}
It follows that a ground state of the X-cube model can be obtained by contracting all the virtual legs:

\begin{equation}
	\ket{\text{X-cube}} = \sum_{\{s_l\}}  \text{tTr}  \left[  \otimes_vT \otimes_l g^{s_l }  \right]  \ket{\{s_l\}}.
\end{equation}
Now we discuss the adaptive protocol that can prepare the above state in constant depth. First, for each vertex, one uses the $T$ tensor to construct a stabilizer state of 6 qubits in the $Z$ basis (Fig.\ref{fig_fracton}a)

\begin{equation}\label{eq:x_cube_vertex}
	\ket{T}_v = \sum_{x,\bar{x}, y,\bar{y},z,\bar{z}}  T_{x\bar{x}, y\bar{y},z\bar{z}} \ket{x,\bar{x}, y,\bar{y},z,\bar{z}  }.
\end{equation}
There are three independent Z-type stabilizers,  each of which is a  product of four Pauli-Zs  for each x-y, y-z, or z-x plane.  There are eight (not all independent) X-type stabilizers, each of which takes the form $XXX$ where each $X$ lives in distinct spatial directions. It is also useful to notice that the product of those X-type stabilizers can generate the two-body X-stabilizers: $X_x X_{\bar{x}}$, $X_y X_{\bar{y}}$, and $X_z X_{\bar{z}}$.  

We first simultaneously prepare the vertex state  $\otimes_v \ket{T}$, and then  perform the two-body ZZ measurements on all two qubits sharing the same link, leading to the following state

\begin{equation}
	\ket{\psi} = \left[\prod_{l} \frac{1+ ZZ}{2}  \right] \otimes_v \ket{T},
\end{equation}
when every measurement outcome is one. Such a measurement effectively ``fuses'' two qubits into a single qubit on every link, thereby  realizing the ground state of the X-cube model.

Now we discuss how to correct the measurement errors ($ZZ=-1$).  We notice that since the product  of all $ZZ$  operators on each plane stabilizes the state $ \otimes_v \ket{T}$, the   errors come in pairs on each plane. This means that there are two following fundamental errors. The first type of error occurs on two links along the same line, which can be annihilated using a product of Pauli-Xs in between (Fig.\ref{fig_fracton}b). The second type of error corresponds to three errors on three lines that intersect at a vertex, and can be corrected by a product of X emitted from the vertex  (Fig.\ref{fig_fracton}c). Physically these measurement errors can be understood as lineon  excitations  in the X-cube model, which can be annihilated by the product of X described above. In particular, the first type of error corresponds to the fact that without extra energy cost, a lineon can only move along a 1d line, whereas the second type of error reflects the fact that a lineon can change direction by paying an energy cost for creating an extra lineon.

\section{Adaptive circuits via disentangling fluctuating domain walls}

\subsection{1d SPT with $\mathbb{Z}_2 \times \mathbb{Z}_2$ symmetry}\label{sec:1d_spt}

On a 1d lattice of 2L lattice sites with periodic boundary conditions, we consider a cluster state Hamiltonian 
\begin{equation}
H_ g= -  \sum_{i=1}^{2L} Z_{i-1}X_i Z_{i+1}.
\end{equation}
The ground state exhibits an SPT order protected by a $\mathbb{Z}_2 \times \mathbb{Z}_2$ symmetry generated by the product of Pauli-Xs on odd sites and even sites respectively.  

Using $ \ket{ \{a_i \}, \{b_{i}\}    }$ to denote the computation basis with $a_i$ defined on odd sites and $b_{i}$ defined on even sites, the ground states can be written as 
\begin{equation}
\ket{\text{SPT}} = \sum_{   \{a_i \}, \{b_{i}\} }  \prod_{i=1}^{L} (-1)^{ b_i(a_i + a_{i+1})   } \ket{ \{a_i\} ,\{b_{i}\}  },
\end{equation}
since it can be obtained from the product state $\ket{ +}^{\otimes 2L}$ by applying control-Z gates acting on every two neighboring qubits. To observe the braiding of fluctuating domain walls, we express the state in the Pauli-X basis: 
\begin{equation}
\ket{\text{SPT}}  =  \sum'_{ \{\alpha_i\},\{\beta_{i}\} }   (-1)^{ \chi(\alpha, \beta )  }   \ket{  \{ \alpha_i  \} , \{\beta_{i}   \}  },
\end{equation}
with $\alpha_i, \beta_i  = \pm 1 $ corresponding to spins on odd and even sites respectively in X basis. Here,  the two $Z_2$ symmetry actions on odd and even  sites impose the constraints $\prod_{i=1}^L \alpha_i= \prod_{i=1}^L\beta_i =1$. Therefore, for any allowed $\{\alpha_i\}$ configuration, $\alpha_i=-1$ must come in pairs, which can be connected by a string. Similarly one can use strings to enumerate all possible $\{\beta_i\}$ configurations. $\chi (\alpha, \beta )$ is the number of times that $\alpha$ braids with $\beta$ configurations.

Now when we measure Pauli-X on every odd site, one obtains a particular $\{\alpha_i\}$ configuration, and crucially, the un-measured degrees of freedom $\beta$ is a superposition of strings, which exhibits a long-range $\mathbb{Z}_2$ order, i.e. $\expval{Z_iZ_j}=\pm 1$ for any $i,j$.

\subsection{Details on preparing CSS codes}\label{appendix:disentangling_css}

\textbf{Chain complex description} \cite{bravyi2014homological,kubica2018ungauging} -- Given the qubits labeled by $q\in B_Q $, any CSS code Hamiltonian takes the form $H=- \sum_{i \in \mathcal{B}_X}  S_i^X  - \sum_{j\in \mathcal{B}_Z} S_j^Z$, where the stabilizer $S_i^X$ and $S_j^Z$ are products of Pauli-X and Z operators respectively. These stabilizers can be expressed via the parity check matrix $H_X$ and $H_Z$: the $i$-th row of $H_X$ corresponds to the support of the X-type stabilizer $S_i^{X}$ with $i \in B_X$, and similarly, the $j$-th row of $H_Z$ corresponds to the support of the Z-type stabilizer $S_j^{Z}$ with $j \in B_Z$, e.g. $S_j^Z = \prod_{q\in B_Q}   Z_q^{ [H_Z]_{jq} } $. Since every stabilizer commutes, the parity check matrix satisfies $H_Z  H_X^T=0$. This is the crucial property that enables the chain complex description of CSS codes by identifying the boundary operators $\partial_X$ and $\partial_Z$ as the parity check matrices, i.e. $H_X^{T}=\partial_X$ and $H_Z =\partial_Z$. The commutativity of stabilizers (i.e. $H_ZH_X^T=0$) then translates to a topological fact that the composition of the boundary maps must be  zero: $\partial_Z\partial_X =0$. Equipped with the boundary maps, the chain complex associated with a CSS code reads

	\begin{equation}
		\includegraphics[width=6.5cm]{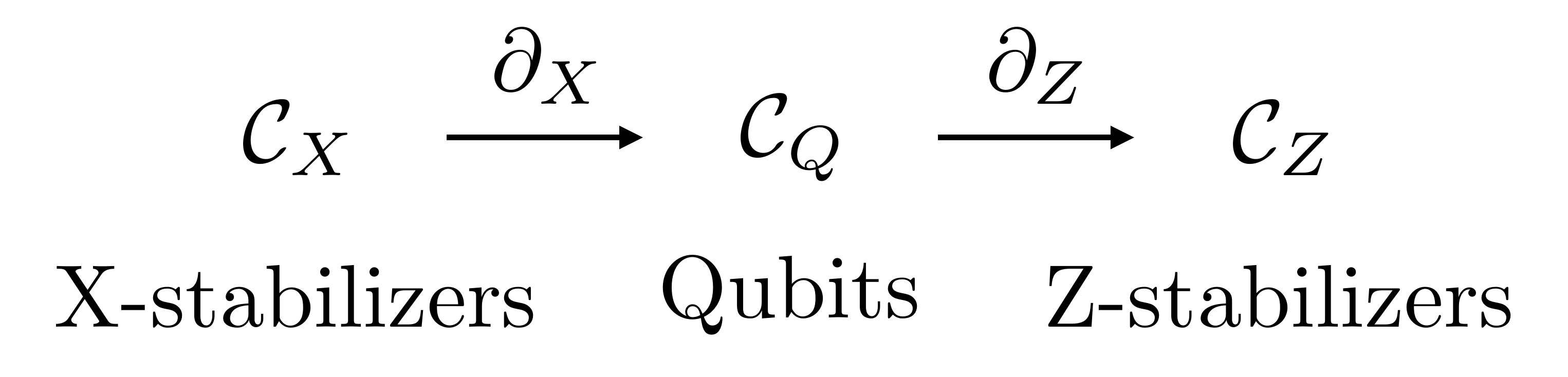}
	\end{equation}
	where $C_X$, $C_Q$ and $C_Z$ are finite-dimensional $\mathcal{F}_2$-vector spaces with bases $\mathcal{B}_X$ = X-type stabilizers, $\mathcal{B}_Q$ = qubits and $\mathcal{B}_Z$ = Z-type stabilizers. Using the identification between boundary maps and the parity check matrices, we write stabilizers as $ S_j^Z = \prod_{q\in B_Q}   Z_q^{ [H_Z]_{jq} }  = \prod_{q\in B_Q}   Z_q^{ [\partial_Z]_{jq} }  $ and $ S_i^X = \prod_{q\in B_Q}   X_q^{ [H_X]_{iq} }  = \prod_{q\in B_Q}   X_q^{ [\partial_X]_{qi} }$, and therefore the CSS code Hamiltonian is 
	
	\begin{equation}
		H=  - \sum_{i \in\mathcal{B}_X}  	 \prod_{q\in B_Q}   X_q^{ [\partial_X]_{qi} }  - \sum_{j \in\mathcal{B}_Z}   \prod_{q\in B_Q}   Z_q^{ [\partial_Z^T]_{qj} }. 
	\end{equation}

It may also be  succinctly expressed as 
	
	\begin{equation}\label{eq:css}
			H=  - \sum_{i \in\mathcal{B}_X}     X(  \partial_X  i  )    - \sum_{j \in\mathcal{B}_Z}  Z(\partial_Z^T j),  
	\end{equation} 
where $ X(  \partial_X  i  ) $ is the product of Pauli-Xs for spins on the boundary of $i\in \mathcal{B}_X$, and $ Z(  \partial_Z^T  j   ) $ is the product of Pauli-Zs for spins on the coboundary of $j\in \mathcal{B}_Z$.

An illuminating example is the 2d toric code, which can be obtained  by associating the chain complex with a 2d lattice. In this case, $C_2$, $C_1$, and $C_0$ are associated with faces (2-cells), edges (1-cells) and vertices (0-cells). Denoting $\mathcal{B}_X$ and $\mathcal{B}_Z$ as the bases for $C_2$ and $C_0$, one finds the toric code Hamiltonian for qubits living on 1-cells: 
	
	\begin{equation}
		H=  - \sum_{i \in\mathcal{B}_X}     XXXX    - \sum_{j \in\mathcal{B}_Z} ZZZZ,
	\end{equation} 
	where the first term is the product of $X$ on the boundary of a 2-cell, and the second term is the product of $Z$ on the coboundary of a 0-cell. \\

\textbf{Preparation protocol} -- Now we construct an SRE state that leads to the CSS code (Eq.\ref{eq:css}) upon measurements. First, in addition to the spins labeled by $q \in \mathcal{B}_Q$, we also introduce spins labeled by $j\in \mathcal{B}_{Z}$. Initializing all spins in the $+1 $ eigenstate of Pauli-X, one introduces an  operator $Z(\partial_Z q )$ that creates a domain-wall configuration for the spins ($\in \mathcal{B}_Z$) located on the boundary of $q$. $Z(\partial_Z q )$ for all $q\in \mathcal{B}_Q$ then forms a generating set to generate all possible domain-wall configurations denoted by $\ket{\alpha}$ for spins localized in $\mathcal{B}_Z$. Similarly, $Z(\partial^T_Z j )$ for all $j\in \mathcal{B}_Z$  forms a generating set to generate all possible dual domain-walls  denoted by $\ket{\beta}$ for spins localized in $\mathcal{B}_Q$.

It follows that braiding two types of fluctuating domain walls leads to the  SRE state

\begin{equation}
\ket{\psi_0} 	 = \sum'_{\alpha, \beta  } (-1 )^{\chi(\alpha,\beta)} \ket{
\alpha,\beta}, 
\end{equation}
where $\chi(\alpha,\beta)$ is a generalized braiding (linking) number. Alternatively, the state can be written as $\ket{\psi_0} = \prod_{q\in \mathcal{B}_Q } (1+X_q Z(\partial_Zq)) \prod_{j\in \mathcal{B}_Z } (1+X_j Z(\partial^T_Z j)) \otimes \ket{+}$. One can also write down a gapped stabilizer Hamiltonian $H_0$ for which $\ket{\psi_0}$ is the ground state: 
\begin{equation}
H_0 = - \sum_{j \in\mathcal{B}_Z}  X_j Z(\partial_Z^T j) -  \sum_{q \in B_Q}  X_q  Z(\partial_Z q) \label{eq:graphstate}
\end{equation}
$X_j Z(\partial_Z^T j)= 1$ enforces the condition that $Z(\partial_Z^T j)$ creates a domain wall on the coboundary of $j$ with a sign that depends on whether $X_j= \pm 1$, and similarly,  $X_{q} Z(\partial_Z q)= 1$ enforces the condition that $Z(\partial_Z q)$ creates a domain wall on the boundary of $q$ with a sign that depends on whether $X_q= \pm 1$. These two  rules naturally lead to an SRE state as two species of fluctuating domain walls with a braiding phase. In fact, Eq.\ref{eq:graphstate} is simply a cluster state Hamiltonian defined on a bipartite graph.

Given the SRE state $\ket{\psi_0}$, measuring Pauli-X  for spins labeled by $j\in \mathcal{B}_Z$ projects those spins to a particular domain-wall configuration $\ket{\alpha'}$, which results in the state

\begin{equation}\label{}
\ket{\alpha'}\otimes  \sum_{\beta}    (-1 )^{\chi(\alpha',\beta)}     \ket{\beta}   \longrightarrow	 \ket{\text{CSS}}  = \sum_{\beta}  \ket{\beta}.
	\end{equation}
Here the arrow denotes the procedure: (i) getting rid of the product state $\ket{\alpha'}$ (ii) removing the sign $(-1)^{\chi(\alpha',
\beta)}$ by applying $X(c_q)$, i.e. a product of Pauli-Xs (acting on qubits in $\mathcal{B}_Q$) where $c_q$ denotes the set of qubits whose boundary generates the $\alpha'$ configuration via $Z(\partial_Z c_q)$.  Formally, $\ket{\text{CSS}}$ can be written as a product state $\prod_{j \in B_Z}   \left[ 1+ Z( \partial_Z^T j ) \right] \otimes_q \ket{+}_q$, which can be expanded as a sum of all possible domain-wall creation operators. This expression also manifests the fact that $\ket{\text{CSS}}$ is a ground state of the CSS Hamiltonian (Eq.\ref{eq:css}).

\textbf{Construction from the level of operators} -- Above we discuss the adaptive preparation of CSS codes in terms of wave functions, here we provide a complementary description in terms of operators. Given the Hamiltonian $H_0$ (Eq.\ref{eq:graphstate}), we measure Pauli-X on spins labeled by $j\in \mathcal{B}_Z$ for the ground state of $H_0$. It follows that  $X_jZ(\partial_Z^T j)$ is replaced by $\pm Z(\partial_Z^T j)$ where the sign is determined by the measurement outcomes. As for the second term $X_qZ(\partial_Z q)$, since $q (\in \mathcal{B}_Q)$ has a boundary belonging to $B_Z$,  $Z(\partial_Z q)$ would not commute with the measurement on $j$-th spins for $j \in \mathcal{B}_Z$. It follows that the measurement will induce new terms that commute with the measurement by taking a product of $X_q  Z(\partial_Z q)$ over various $q\in \mathcal{B}_Q$. As a result we are looking for a cycle $c = \sum_{\{c_q \in {0,1} \}} c_q q\in C_Q$ so that $c$ has no boundary (i.e. $\partial_Z c =0$ or equivalently, $c\in \text{Ker}(\partial_Z)$). The corresponding generated term is $\left[\prod_{q \in \mathcal{B}_Q   } X_q^{c_q}   \right]  Z( \partial_{Z} c   )  =  \prod_{q \in \mathcal{B}_Q   } X_q^{c_q} $. 

To conclude, given the graph state Hamiltonian Eq.\ref{eq:graphstate}, measuring Pauli-X for spins defined on $j  \in \mathcal{B}_Z$ gives the post-measurement Hamiltonian that describes the spins labeled by $q  \in  \mathcal{B}_q$:

\begin{equation}
H=        \sum_{j \in\mathcal{B}_Z} \pm   Z(\partial_Z^T j)  -    \sum_{ c \in \text{Ker}(\partial_Z)}  ~  \prod_{q \in \mathcal{B}_Q   } X_q^{c_q}.
\end{equation}
As $\partial_Z\partial_X =0$, any $c$ that belongs to the image of $\partial_X$ must have $\partial_Z=0$, this means one choice of $c$ would be $c= \partial_X i $ for $i \in \mathcal{B}_X$. On the other hand, there can be homologically nontrivial cycle $c$ not in the image of $\partial_X$, meaning $c \in \text{Ker}(\partial_Z)/ \text{Imag}(\partial_X)$. Therefore, the post-measurement Hamiltonian can be written as $H=$

\begin{equation}	
     \sum_{j \in\mathcal{B}_Z}\pm  Z(\partial_Z^T j)  -    \sum_{ i \in \mathcal{B}_X } X(\partial_X i) - \sum_{c \in \text{Ker}(\partial_Z)/ \text{Imag}(\partial_X) }   X(c).
\end{equation}
The first and the second term exactly reproduces the desired CSS code Hamiltonian Eq.\ref{eq:css} (up to a sign that can be corrected by a depth-1 unitary circuit), and the last term simply gives the logical operators that specify a unique state in the ground state subspace.

\section{Operator pushing in symmetric subspace}\label{appendix:spt_push}
Here we provide details on operator pushing and its generalization in the presence of symmetry. The statement of  the operator pushing is the following theorem: 

\textbf{Theorem}: given  a bipartite pure state $\ket{\psi } \in \mathcal{H}_A \otimes \mathcal{H}_B$ where $A$ and $B$ have equal Hilbert space dimension $d$, for any unitary $U_A$ supported on $A$, there always exists a unitary $V_B$ supported on $B$ such that  $U_A\ket{\psi} =V_B \ket{\psi}$, if and only if $\ket{\psi}$ is maximally entangled between $A$ and $B$ (i.e the entanglement entropy is $\log d$).

A motivating example is given by a Bell pair shared between two qubits: $\ket{\text{Bell}} = \frac{1}{\sqrt{2}} \left( \ket{00} +\ket{11}\right)$. Since $X_AX_B \ket{\text{Bell}}   = Z_AZ_B \ket{\text{Bell}}$ (equivalently, $\ket{\text{Bell}}$ is a stabilizer state stabilized by $X_AX_B$ and $Z_AZ_B$), one finds $X_A\ket{\text{Bell}} = X_B \ket{\text{Bell}}$, $Z_A \ket{\text{Bell}}  = Z_B \ket{\text{Bell}}$, and $Z_AX_A \ket{\text{Bell}} = X_BZ_B \ket{\text{Bell}}$. Since any unitary operator $U_A$ can be expanded in the operator basis as $U_A= \sum_{i,j \in\{0,1\}} u_{ij} X^i_A Z_A^j$, its action on $\ket{\text{Bell}}$ is equivalent to $V_B= \sum_{i,j \in\{0,1\}} u_{ij}  Z_B^jX^i_B$, whose unitarity inherits from $U_A$.

\textbf{Proof sketch} (see also Ref.\cite{cresswell_2020_pusing}): to start,  one writes any state $\ket{\psi} $ in its schmidt decomposition $\ket{\psi}= \sum_{i=1}^{d}  \sigma_i \ket{i}_A\ket{i}_B$ with $\sigma_i\geq 0$ and $\sum_{i=1}^d \sigma_i^2 = 1$. Applying an operator $U_A$ supported on $A$ gives

\begin{equation}
U_A \ket{\psi}	=  \sum_{i,j =1}^d \sigma_i   U_{ji}   \ket{j}_A \ket{i}_B.  
\end{equation}  
On the other hand, applying  an operator $V_B$ supported on $B$ gives

\begin{equation}
V_B \ket{\psi}	=  \sum_{i,j =1}^d \sigma_i  V_{ji}   \ket{i}_A \ket{j}_B. 
\end{equation} 
These two actions are equivalent if and only if $U\Sigma = \Sigma V^T$ where $\Sigma$ is a diagonal matrix with $\{\sigma_i\}$ being the diagonal entries. When $\ket{\psi}$ is maximally entangled, $\Sigma$ is proportional to an identity matrix, in which case, the unitary $U$ on $A$ corresponds to a unitary $V= U^T $ on $B$. On the other hand, demanding the unitarity $V^{\dagger} V =\mathbb{I}$ for all $V= \Sigma U^T \Sigma^{-1}$ implies $[U,\Sigma^2]=0$ for all $U$. This in turns indicates $\Sigma\propto \mathbb{I}$.

\textbf{Operator pushing with symmetry}: given two Hilbert spaces  $\mathcal{H}_A, \mathcal{H}_B$, which may have unequal dimensions, we impose symmetry $S_A$ supported on $A$ and $S_B$ supported on $B$ to define the symmetric subspaces $H_{A,\text{symm}} $ and $H_{B,\text{symm}} $. In other words, any  state $\ket{\psi_{\text{symm}}}$ in  $H_{A,\text{symm}} \otimes H_{B,\text{symm}}$ is invariant under the symmetry action by $S_A$ and $S_B$. For the case $\text{dim}(\mathcal{H}_{A,\text{symm}}) =  \text{dim}(\mathcal{H}_{B,\text{symm}}) = d_{\text{symm}}$, we consider a symmetric state $\ket{\psi_{\text{symm}}} $ with maximal entanglement  in the symmetric subspace, i.e. the entanglement entropy between $A$ and $B$ is $\log d_{\text{symm}}$. We now consider applying a symmetric unitary  $U_{A,\text{symm}}$ acting on $A$. Due to the symmetry, $U_{A,\text{symm}}$ remains a unitary when restricted in the symmetric subspace. Therefore, one can directly apply the aforementioned theorem to find that, for  $U_{A,\text{symm}}$ acting on $A$, there exists a symmetric unitary  $V_{B,\text{symm}}$ acting on $B$ so that $U_{A,\text{symm}}\ket{\psi_{\text{symm}}} =V_{B,\text{symm}} \ket{\psi_{\text{symm}}}  $. Namely, one can push any symmetric unitary from $A$ to $B$ through the maximally entangled state in the symmetric subspace.

\section{Bound on entanglement growth}\label{appendix:entanglement_bound}

Here we provide a bound on entanglement generated by local adaptive quantum circuits. Specifically, consider a depth-$D$ adaptive quantum circuit, where each layer consists of non-overlapping local gates that can be a unitary or a projector due to measurement, we show that the increase of entanglement of a subsystem $A$ is bounded by $O(D|\partial A|)$ with $|\partial A|$ being the boundary area of $A$. Here the entanglement for a bipartite pure state $\ket{\psi}_{AB} \in \mathcal{H}_A\otimes \mathcal{H}_B$ is quantified by the max-entropy (also dubbed Hartley entropy) $S_0= \log \chi$, where $\chi$ is the rank of the subregion reduced density matrix $\rho_A= \tr_B \ket{\psi}\bra{\psi}$. Equivalently, $\chi$ is the Schmidt rank in the Schmidt decomposition  $|\psi\rangle_{AB}  = \sum_{i=1}^\chi \sqrt{p_i}\ket{\phi_i}_A\otimes |\theta_{i}\rangle_B$ with $\{ |\phi_i \rangle_A \}$ and $\{|  \theta_i   \rangle_B \}$ being the orthonormal basis sets for $A$ and $B$ respectively.

Note that while the above result only bounds the increase of max-entropy, one can immediately obtain the following statement: given a  product state, the output state of a depth-$D$ adaptive quantum circuit can at most have $O(D|\partial A|)$ Renyi entanglement entropy $S_n=\frac{1}{1-n} \log \left(  \tr_A \rho_A^n \right)$ for any positive $n$ (including the von-Neumann entropy by the limit $n\to 1$). This is because our bound on entanglement growth implies that the output state can at most has $O(D|\partial A|)$ max-entropy $S_0$, and $S_0 \geq S_n$ for any $n>0$.

As a non-trivial application, the result above implies starting from a product state, the preparation of  a 1d CFT critical state of size $L$ via local adaptive circuits requires a depth $D\gtrsim O(\log L)$ due to the logarithmic scaling of entanglement in a 1+1D CFT. Therefore, the $(\log L)$-depth adaptive circuit we introduce in Sec.\ref{sec:mera} for preparing a critical state is optimal.

Below we present our proof for bounding the entanglement growth of max-entropy. The proof relies on the following two facts :
\vspace{2mm}

\noindent(i) \textit{A gate acting only on $A$ or $B$ cannot increase the entanglement max-entropy.} To see this, one can  consider  a gate $O_A$ supported on region $A$. After the application of $O_A$, a state $\ket{\psi}_{AB}$ in its Schmidt decomposition  $|\psi\rangle_{AB}  = \sum_{i=1}^\chi \sqrt{p_i}\ket{\phi_i}_A\otimes |\theta_{i}\rangle_B$ will transform to
\begin{equation}
	(O_A \otimes I_B)|\psi\rangle_{AB} = \sum_{i=1}^{\chi} \sqrt{p_i} O_A|\phi_i\rangle_A \otimes | \theta_i\rangle_B.
\end{equation}
Under the bijective map $|\theta_i\rangle_B \leftrightarrow \langle \theta_i|_B$,  the state above can be viewed as a matrix of rank at most $\chi$. By performing the singular value decomposition of this matrix, one can obtain the Schmidt decomposition of $(O_A \otimes I_B)|\psi\rangle_{AB}$. Since the rank of the matrix is at most $\chi$, the Schmidt rank of $O_A \otimes I_B |\psi\rangle_{AB}$ is at most $\chi$ as well. The same reasoning applies to an operator acting on $B$. \\

\noindent(ii) \textit{A gate $O_{AB}$ acting on the bipartition boundary of $O(1)$ size (i.e. it acts on both $A$ and $B$) can at most increase  entanglement max-entropy by an $O(1)$ value}. To see this, one can apply a Schmidt decomposition to express  $O_{AB}$ as 

\begin{equation}
O_{AB}  = \sum_{\alpha=1}^{\chi' }  \lambda_{\alpha} O_{A,\alpha} \otimes O_{B,\alpha},  
\end{equation}
with $O_{A,\alpha}, O_{B,\alpha}$ supported on $A, B$ respectively. Applying this operators on $\ket{\psi}_{AB} $ gives
\begin{equation}
\sum_{\alpha=1}^{\chi'}\sum_{i=1}^{\chi} \lambda_{\alpha} \sqrt{p_i} \left( O_{A,\alpha}|\phi_i\rangle_A \right)\otimes \left(O_{B,\alpha}| \theta_i\rangle_B \right).
\end{equation}
Clearly, the Schmidt rank of the state above is at most $\chi'\chi$, meaning the increase of the bipartite max-entropy  $S_0$ is upper bounded by $\log \chi'$. Since the operator $O_{AB}$ acts on a region of $O(1)$ size, $\chi'$ is bounded by an $O(1)$ number, and so is $\log \chi'$.

Combining the above two facts leads to the conclusion that the increase of  entanglement max-entropy is  bounded by the number of boundary gates in a depth-$D$ adaptive circuit, independent of the measurement outcomes. Since each layer contains $O(\partial A)$ boundary gates and there are $D$ layers, one finds the growth of entanglement max-entropy is upper bounded by $O(D|\partial A|)$.

Finally, we remark that it is tempting to argue the increase of von-Neumann entanglement entropy obeys the same area-law bound, but this is generally false. Consider, for example, a superposition of the trivial state and a Haar random state, with amplitudes $\sqrt{1-\epsilon}$ and $\sqrt{\epsilon}$ respectively. Certainly, by choosing a sufficiently small $\epsilon$, one can ensure that von-Neumann entanglement entropy for any bipartition takes an  $O(1)$ value. However, even a local measurement can lead to a volume-law entanglement growth by post-selecting on the Haar-random state. Such an example therefore invalidates an area-law bound.

\end{document}